\documentclass[apj]{emulateapj}
\usepackage{savesym}
\savesymbol{tablenum}
\usepackage{siunitx}
\restoresymbol{SIX}{tablenum}

\slugcomment{Submitted to ApJ February 2016}
\usepackage[innercaption]{sidecap}
\sidecaptionvpos{figure}{m}

\shorttitle{Transitions in the cloud composition of hot Jupiters}
\shortauthors{Parmentier et al.}
\usepackage{natbib}
\usepackage{subfigure,wrapfig}
\usepackage[]{sidecap}
\usepackage[T1]{fontenc}
\usepackage{mathptmx}
\usepackage{hyperref}

\hypersetup{
   colorlinks=true,
   citecolor=blue,
   linkcolor=blue,
   urlcolor=black
   }
   
   \usepackage[figuresright]{rotating}

\def\ct{}

\begin{document}

\title{Transitions in the cloud composition of hot Jupiters}

\author{Vivien Parmentier\altaffilmark{1,2,4}, Jonathan J. Fortney\altaffilmark{1}, Adam P. Showman\altaffilmark{2}, Caroline Morley\altaffilmark{1}, Mark S. Marley\altaffilmark{3}}

\altaffiltext{1}{Department of Astronomy and Astrophysics, University of California, Santa Cruz, CA 95064}
\altaffiltext{2}{Department of Planetary Sciences and Lunar and Planetary Laboratory, The University of Arizona, Tucson, AZ 85721, USA}
\altaffiltext{3}{NASA Ames Research Center, MS-245-3, Moffett Field, CA 94035, USA}
\altaffiltext{4}{NASA Sagan Fellow}



\begin{abstract}
Over a large range of equilibrium temperatures, clouds shape the transmission spectrum of hot Jupiter atmospheres, yet their composition remains unknown. Recent observations show that the \emph{Kepler} lightcurves of some hot Jupiters are asymmetric: for the hottest planets, the lightcurve peaks before secondary eclipse, whereas for planets cooler than $\sim1900\rm\,K$, it peaks after secondary eclipse. We use the thermal structure from 3D global circulation models to determine the expected cloud distribution and \emph{Kepler} lightcurves of hot Jupiters. We demonstrate that the change from an optical lightcurve dominated by thermal emission to one dominated by scattering (reflection) naturally explains the observed trend from negative to positive offset. For the cool planets the presence of an asymmetry in the \emph{Kepler} lightcurve is a telltale sign of the cloud composition, because each cloud species can produce an offset only over a narrow range of effective temperatures. By comparing our models and the observations, we show that the cloud composition  of hot Jupiters likely varies with equilibrium temperature. We suggest that a transition occurs between silicate and manganese sulfide clouds at a temperature near $1600\rm\,K$, analogous to the L/T transition on brown dwarfs. The cold trapping of cloud species below the photosphere naturally produces such a transition and predicts similar transitions for other condensates, including TiO. We predict that most hot Jupiters should have cloudy nightsides, that partial cloudiness should be common at the limb and that the dayside hot spot should often be cloud-free.

 
\end{abstract}


\section{Introduction}

Clouds play a major role in shaping the transmission spectra of hot Jupiters, impeding a precise determination of molecular abundances~\citep{Sing2016}. Despite their importance, the composition of these clouds remains unknown, partly because their optical properties are more sensitive to physical properties such as the particle size than chemical composition~\citep{Heng2013} and partly because their vertical distribution is very sensitive to unconstrained processes such as vertical mixing~\citep{Morley2013,Lee2015}. To date no correlation has been observed between the presence of clouds and fundamental planetary parameters such as irradiation or gravity although it is theoretically expected~\citep{Sudarsky2000}. This is surprising since over the same range of effective temperatures a strong correlation between temperature and cloudiness has been established for brown dwarfs~\citep{Kirkpatrick2005,Marley2010}. Clouds appear in the M/L transition and disappear at the L/T transition, during which some objects harbor patchy cloudiness~\citep{Crossfield2014}. 

\begin{figure}[htb] 
 \centering
  \includegraphics[width=\linewidth]{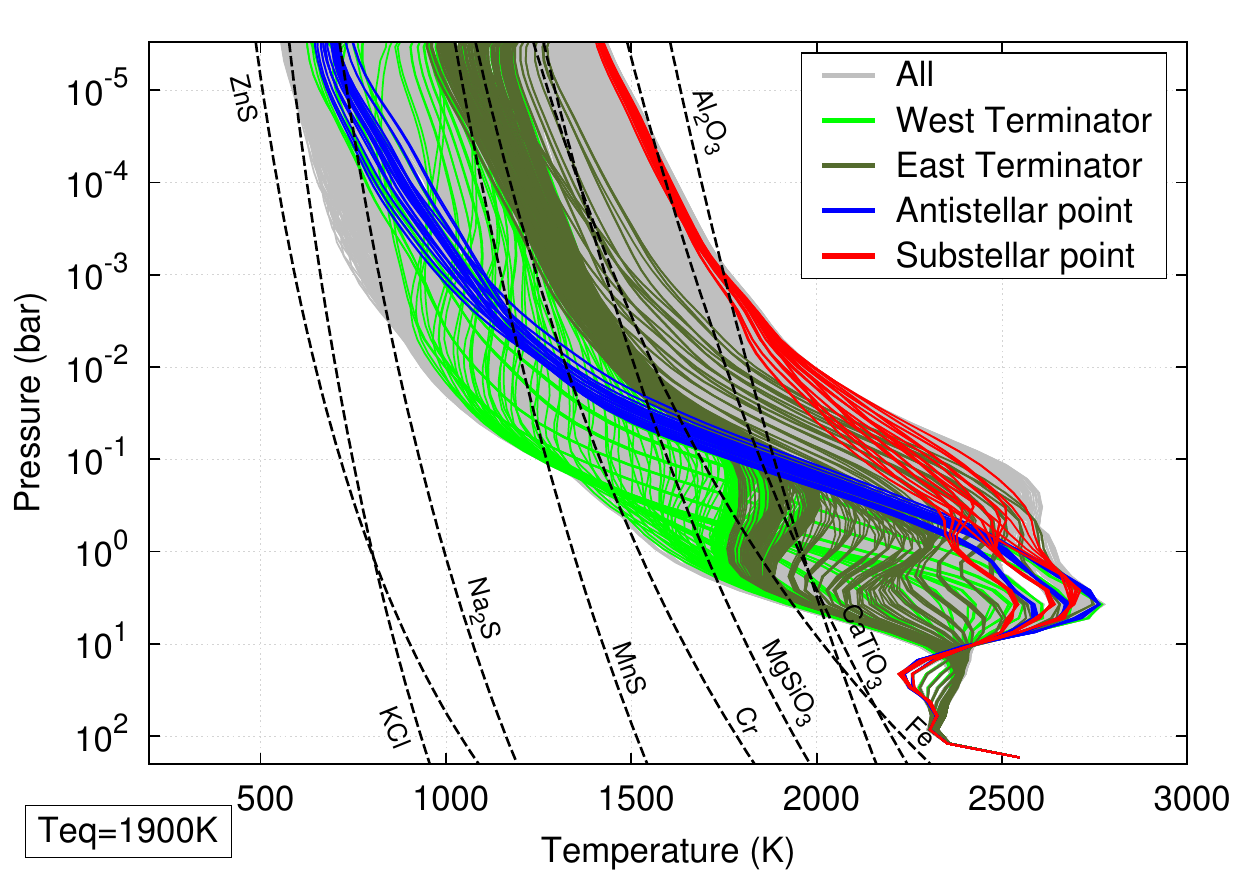}
  \caption{ Pressure temperature profiles from our fiducial model for a planet with $T_{\rm eq}=1900\,K$ (grey: profiles at all latitudes and longitudes, light green: profiles at the limb west of the substellar point, dark green:  profiles at the limb east of the substellar point, blue: profiles at less than $20^{\circ}$ of the antistellar point, red: profiles at less than $20^{\circ}$ of the substellar point). The condensation curves of several important species are plotted as dashed lines. At a given pressure, if the temperature is cooler than the condensation temperature, we consider the atmosphere cloudy. For pressures lower than $P_{\rm top}$, we assume that clouds are not present, with this cloud top pressure used as a free parameter.}
  \label{fig::PT}
\end{figure}

Hot Jupiter atmospheres possess large horizontal temperature contrasts despite the strong circulation driven by the intense and inhomogeneous irradiation they receive. Fast eastward winds and planetary scale waves shift the temperatures eastward: west of the substellar point the atmosphere is cooler than east of it~\citep{Showman2002,Knutson2007}. Different gases are expected to condense at different locations on the planet, leading to a possibly complex, inhomegeneous and asymmetric cloud distribution (see Figure~\ref{fig::PT}). 

\begin{figure}
    \includegraphics[width=\linewidth]{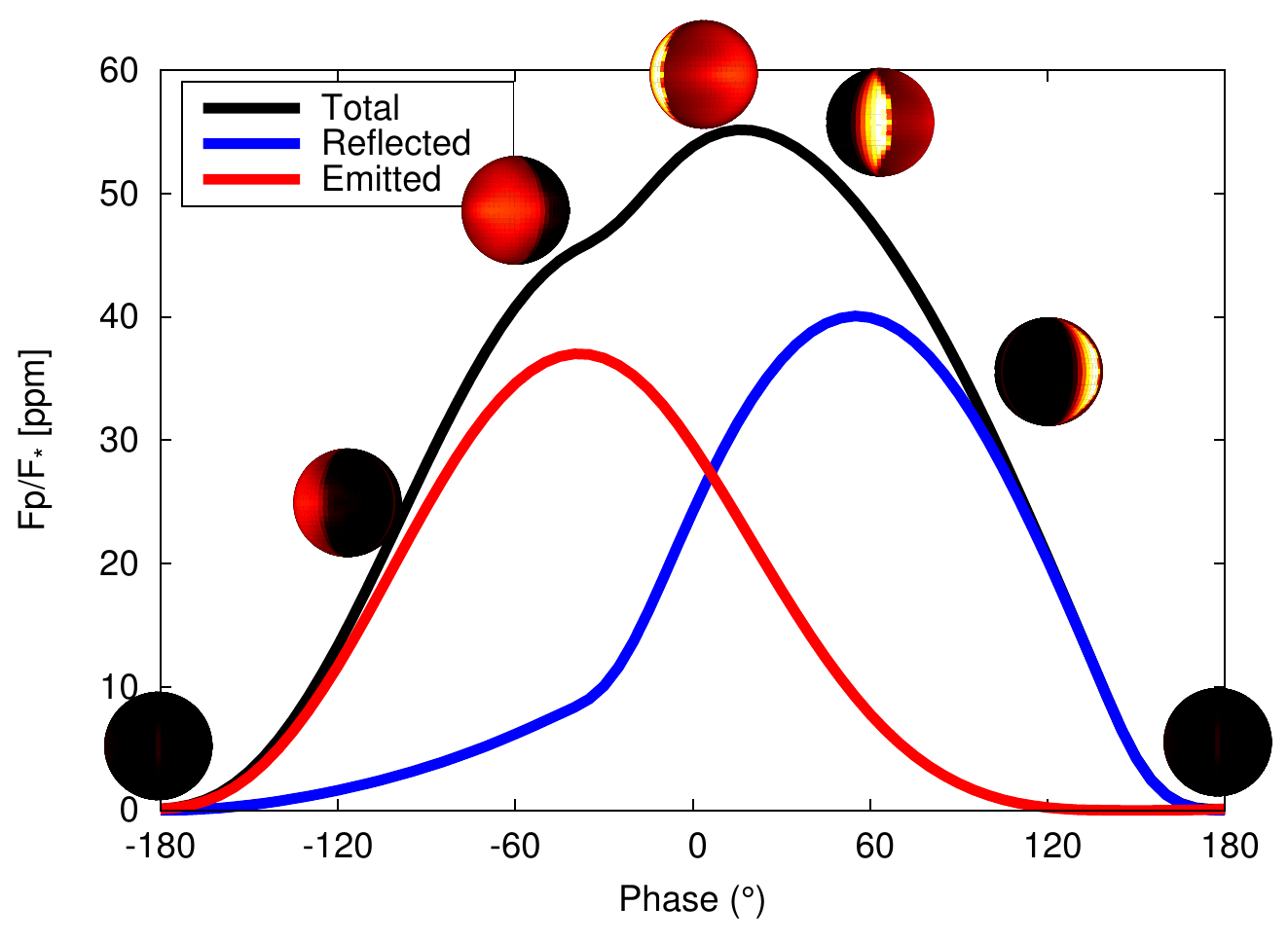}
\caption{Phase curves from our fiducial model for a planet with $T_{\rm eq}=1900\,\rm K$ in the \emph{Kepler} bandpass (black) with the contributions from thermal emission (red) and reflected light (blue). The flux from the planet's visible hemisphere at several phases is also depicted. Secondary eclipse happens at phase 0 and transit at phase $\pm180$. This is for silicate clouds with $a=\SI{0.1}{\micro\meter}$ and $p_{\rm top}=\SI{1}{\micro\bar}$.}
 \label{fig::PC}
\end{figure}

The horizontal structure of the atmosphere can be probed by monitoring the light received from the planet as a function of time. Planetary phase curves are the sum of two contributions: the thermal emission from the planet and the reflected stellar light. The thermal contribution is directly linked to the horizontal temperature variations: the atmosphere emits more light where it is hotter. Hot Jupiters with their eastward shift of the temperature maximum have thermal lightcurves peaking before secondary eclipse (a positive offset). The reflected component provides information on the cloud distribution on the dayside of the planet. Abundant clouds in the coldest part of the dayside atmosphere, west of the substellar point, would lead to a reflected lightcurve that peaks after secondary eclipse (a negative offset)~(see Figure~\ref{fig::PC}). The relative roles of emission and reflection depend both on the planet temperature and the observation bandpass. The \emph{Spitzer Space Telescope} observed the emission-dominated phase curve of several hot Jupiters and confirmed the expected eastward shift of the temperature distribution for a large range of equilibrium temperatures~\citep[see][for a review]{Crossfield2015}.

\begin{figure}
 \centering
  \includegraphics[width=\linewidth]{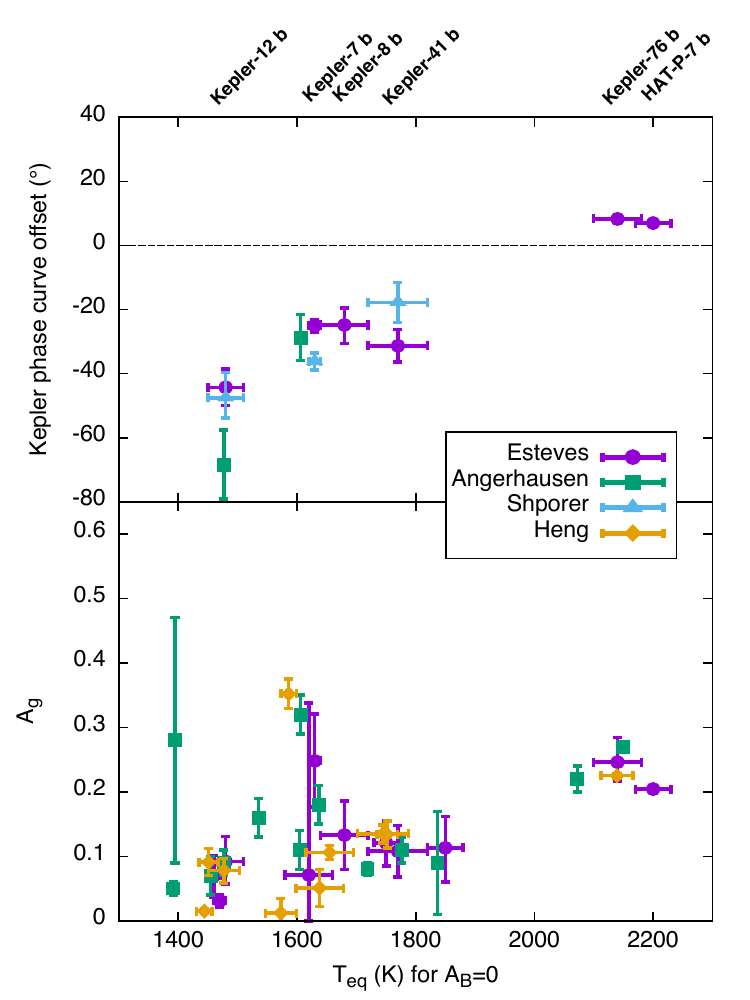}
\caption{
Phase shift of the maximum of the Kepler lightcurve relative to the secondary eclipse (top) and apparent geometric albedo of the planet in the \emph{Kepler} bandpass (bottom) as a function of the equilibrium temperature of the planet. The apparent geometric albedo includes the contribution from both the thermal and the reflected light. Different colors are analyses from different authors. Purple circles are the data from~\cite{Esteves2015} -- used in this study --, green squares are from~\cite{Angerhausen2015}, blue triangles are from~\cite{Shporer2015} and orange diamonds are from~\cite{Heng2013}. Kepler-43b was removed from the analysis as it is doubtful whether the signal is of planetary origin or due to stellar activity~\citep{Esteves2015}}
 \label{fig::Data}
\end{figure}

Recently, shifts in the phase curve of several hot Jupiter systems have been observed in the optical {\ct (from $400$ to $\SI{900}{\micro\meter}$)} by the \emph{Kepler} spacecraft~\citep[e.g.][]{Demory2013}. For the coolest four planets, the lightcurve peaks after secondary eclipse whereas for the hottest two the lightcurve peaks before secondary eclipse, providing the first correlation between equilibrium temperature and cloud coverage for hot Jupiter atmospheres~\citep{Esteves2015,Shporer2015,Angerhausen2015}. Importantly, this trend does not depend on the method used to analyze the data, the number of \emph{Kepler} quarters available or the choice of stellar parameters (see Figure~\ref{fig::Data}). In this paper we choose the values published by~\cite{Esteves2015} but our main conclusions are unaffected by this choice.

In order to understand this correlation we calculate a-priori the thermal structure of a range of cloudless, solar-composition hot Jupiter atmospheres with different equilibrium temperatures using the three-dimensional global circulation model SPARC/MITgcm~\citep{Showman2009}. We then use this thermal structure to determine the longitudinal and latitudinal distribution of clouds in each modeled planet. We assume local equilibrium clouds, meaning that in a given atmospheric cell, all the condensable material condenses until the partial pressure of the remaining gas matches the saturation pressure (see Figure~\ref{fig::PT}). We do not calculate the dynamical mixing of the clouds as done in~\citet{Parmentier2013} nor calculate the particle size distribution as done in~\cite{Ackerman2001}. Instead, we parametrize the vertical mixing and the microphysics by a cloud top pressure below which clouds do not form, and a particle size, both considered as free parameters. Lastly, given the thermal structure and the cloud structure, we model both the thermal and the scattered light from the planet, calculate the phase curve in the \emph{Kepler} bandpass, and compare it to the observations.

Unlike previous work~\citep{Demory2013,Hu2015,Webber2015,Shporer2015,Munoz2015} that aimed to fit ad-hoc cloud models to the Kepler lightcurves by treating the condensation curve of the cloud, the thermal structure of the planet, or the optical properties of the clouds as free parameters, we calculate a-priori the 3D thermal structure and use the condensation temperature and cloud optical properties from known potential condensates to constrain the cloud physical properties and composition. Our work is also different from~\citet{Oreshenko2016} who also compared the temperature map from a global circulation model to the condensation curve of different cloud species.  We aim to for a more detailed calculation of observable signatures, and a much wider comparison to observations.  Our global circulation model uses the full gaseous opacities and not a double-grey framework.  We make no assumptions on the global mixing of the clouds in the planet, and furthermore we use the opacity specific to each cloud species to calculate the phase curves and we compare our models to all known shifted \emph{Kepler} lightcurves, whereas these authors focused on Kepler-7b. As a result we are able to draw conclusions that~\citet{Oreshenko2016} could not, within their study.

Section 2 of this paper describes our suite of models. In Section 3 we consider models with only one cloud species and show that for a given planet, the offset in the \emph{Kepler} lightcurve is determined primarily by the condensation curve of the cloud species rather than the particle size or the vertical mixing. In Section 4, we show that a model where all cloud species are present cannot reproduce the current dataset. We build a physically motivated model where cloud species are removed from the atmosphere when the cloud deck is in the deep atmosphere. This model reproduces simultaneously the albedo and the phase offset of currently observed planets. It implies the disappearance of silicate clouds at equilibrium temperatures lower than $1600\,\rm K$. Finally, we show the cloudiness predicted by our models on the dayside, at the limb and on the nightside of hot Jupiters and discuss the implications for the interpretation of transmission and emission spectra.

\section{Modeling approach}
\subsection{Global circulation model}
To calculate the thermal structure of the planet we solve the global, three-dimensional primitive equations in spherical geometry using the MITgcm, a general circulation model for atmosphere and oceans developed and maintained at the Massachusetts Institute of Technology. We discretize the equations on the cubed-sphere grid as described in~\cite{Adcroft2004} and use a horizontal fourth-order Shapiro filter in order to smooth horizontal noise~\citep{Shapiro1970}. This model has been successfully used to model hot Jupiters atmospheres over the last decade~\citep[e.g.][]{Showman2009,Showman2013,Showman2015,Lewis2010,Lewis2014,Parmentier2013,Kataria2013,Kataria2015}.

We consider Jupiter-sized, tidally locked planets orbiting a solar-type star so that the equilibrium temperature sets the distance and the rotation period of the planet. We use a gravity of $\rm10\rm\,m\,s^{-2}$. The equation of state is ideal gas. We fix $c_p=1.3\times10^4\rm\,J\,kg^{-1}\,K^{-1}$ and use $\kappa=2/7$, appropriate to a predominantly hydrogen atmosphere. The pressure ranges from $\SI{200}{\bar}$ to $\SI{2}{\micro\bar}$ over 53 levels so that we have a resolution of almost 3 levels per scale height. We use a horizontal resolution of C32, equivalent to an approximate resolution of 128 cells in longitude and 64 in latitude and a timestep of $25$ seconds. We initialize the model at rest with a temperature profile from the analytical model of~\cite{Parmentier2015} that uses the analytical expression of~\cite{Parmentier2014a} fitted to represent the global average temperature profile of solar-composition atmospheres without TiO/VO~\citep{Fortney2007}. The simulations all run for 150 earth days, a length over which the observable atmosphere have reached a quasi steady-state~\citep{Showman2009}. Then the first hundred days of the simulation are discarded and the PT profiles are averaged over the last 50 days of simulation.

Titanium and vanadium oxides are trace species that have a dramatic effect on the thermal equilibrium of planetary atmospheres. Despite several attempts to search for TiO and VO in hot Jupiters atmospheres~\citep{Desert2009,Sing2013} weak evidence of its presence has been found only in the hottest known planet~\citep{Haynes2015}. In our fiducial model we considered that TiO and VO have been removed from the atmosphere. The effect of adding TiO/VO opacities in the calculations is investigated in more detail in Section~\ref{Sec::TiO}.

Clouds are expected to have a strong and complex influence in the atmospheric circulation, however given the large number of possible cloud species, their unknown particle size distribution and spatial distribution, a thorough examination of the cloud feedback in the circulation is beyond the scope of this paper. The cloud feedback on the circulation modifies the thermal structure and thus the cloud distribution itself and the phase curve shift. However, as demonstrated in Section~\ref{sec::cloudsFB} for a given cloud setup, this is of second order compared to the effect of the irradiation and does not affect the conclusions based on our cloudless global circulation models.

\subsection{Radiative transfer model}
Radiative transfer is handled both in the 3D simulations and during the post-processing using the plane-parallel radiative transfer code of \cite{Marley1999}. The code was first developed for Titan's atmosphere \citep{McKay1989} and since then has been extensively used for the study of giant planets \citep{Marley1996}, brown dwarfs \citep{Marley2002,Burrows1997}, and hot Jupiters \citep{Fortney2005, Fortney2008, Showman2009}. We use the opacities described in \cite{Freedman2008}, including more recent updates \citep{Freedman2014}, and the molecular abundances described by \cite{Lodders2002a} and \cite{Visscher2006}. 

The version of the code we employ solves the radiative transfer equation in the two-stream approximation using the delta-discrete ordinates method of \cite{Toon1989} for the incident stellar radiation and the two-stream source function method, also of \cite{Toon1989}, for the thermal radiative transfer. Molecular and atomic opacities are treated using  the correlated-k method \citep{Goody1989}: the spectral dimension is divided into a number of bins and within each bin the information of typically 10,000 to 100,000 frequency points is compressed inside a single cumulative distribution function that is then interpolated using 8 $k$-coefficients. For the gas, Rayleigh scattering is taken into account in the calculation. For the clouds, the absorption opacity, the single scattering albedo and the asymmetry parameter are determined with the Mie theory (see Section~\ref{Sec::Clouds}).

To calculate the phase curves we solve the two-stream radiative transfer equations along the line of sight for each atmospheric column and for each planetary phase considering both absorption, emission and scattering.  This method, similar to the calculation of~\citet{Fortney2006a} naturally takes into account geometrical effects such as limb darkening. The stellar flux is assumed to be a collimated flux propagating in each atmospheric column with an angle equal to the angle between the local vertical and the direction of the star. We use 196 frequency bins ranging from $0.26$ to $\SI{300}{\micro\metre}$ and integrate the resulting outgoing flux over the Kepler bandpass.

When coupled to the GCM, the radiative transfer model runs over 11 frequency bins that have been carefully chosen to maximize the accuracy and the speed of the calculation, further details are available in~\citet{Showman2009},~\citet{Parmentier2013} and~\citet{Kataria2013}.

A particular problem appears in the post-processing, when we want to calculate the flux emerging from the planet toward the line of sight within the two-stream approximation. Here we use the two-stream model to compute easily a large number of lightcurve for a given thermal structure. The two-stream model solves the radiative transfer equations for the hemispherically averaged radiation field and is unable to provide information regarding the direction of the radiation escaping a given atmospheric column, contrary to more sophisticated models~\citep[e.g.][]{Cahoy2010,Webber2015}. We assume that the radiation escapes isotropically from the top of each atmospheric column. Asymmetric scattering, however, is considered while solving the two-stream equations and can produce an asymmetric outgoing radiation field. In order to test the validity of the isotropic assumption we use a set of idealized atmospheres characterized by their values of the single scattering albedo $\omega_{0}\in[0,1]$ and their asymmetry parameter $g_{0}\in[0,1]$. For $g_{0}=0$ the atmosphere scatters the light isotropically whereas for $g_{0}=1$ the atmosphere scatters all the photons it received in the forward direction.

\begin{figure}[htb] 
 \centering
  \includegraphics[width=\linewidth]{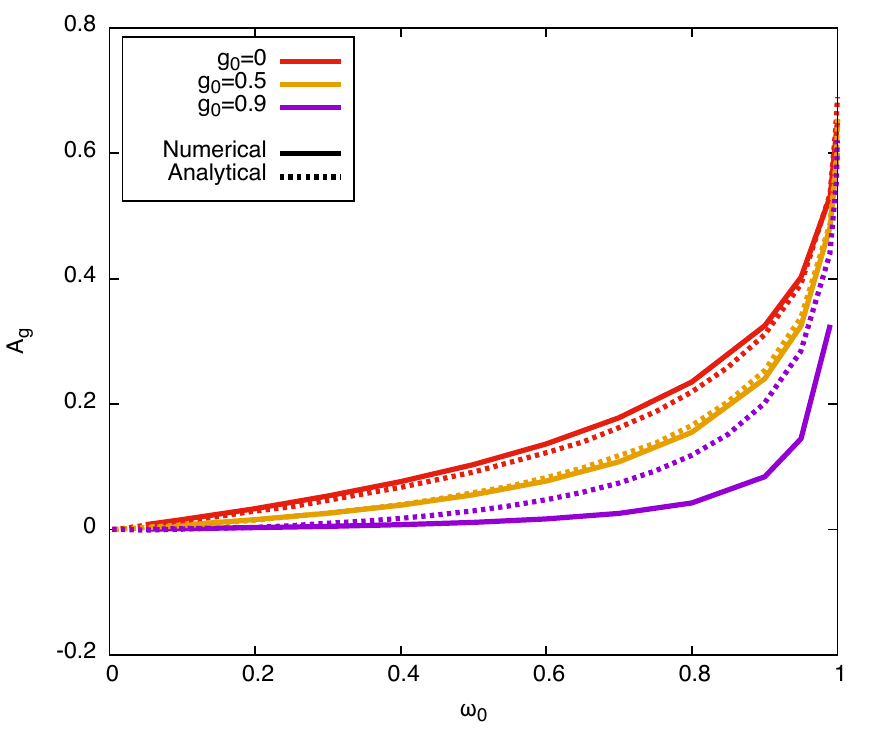}
\caption{
Planet geometric albedo as a function of single-scattering albedo for different asymmetry parameters as calculated by our idealized model (plain lines) and with the analytical solution of~\citet{Madhusudhan2012} (dashed lines). }
 \label{fig::AlbComp}
\end{figure}

In Figure~\ref{fig::AlbComp} we compare our calculation of the geometric albedo calculated with the analytical solution of~\citet{Madhusudhan2012}. Our two-stream scheme with the assumption of isotropic outgoing radiation reproduces well the analytical solution in the small and medium $g_0$ cases but underestimates the albedo in the very forward scattering case ($g_{0}=0.9$). Such a break down for highly forward scattering atmosphere is expected since high values of $g_{0}$ leads to very non-isotropic outgoing flux. The cloud models used in this study have values of the asymmetry parameter close to $0.5$ in the \emph{Kepler} bandpass and are therefore properly represented by our radiative transfer scheme.

\subsection{Cloud model}
\label{Sec::Clouds}

{\ct Numerous condensates can potentially form in the range of temperatures spanned by the the atmospheres of hot Jupiters~\citep{Burrows1999}. There is debate, however, about which condensates will actually form. Numerous mechanisms can promote or inhibit the formation of a given cloud species, leading to a final condensate composition that is not the one predicted by thermochemical equilibrium calculations~\citep[see][for reviews]{Helling2008,Marley2013}. Two main approaches have been used to determine which condensates form in substellar atmospheres. In the~\emph{equilibrium cloud} approach it is assumed that the change of the physical properties of a parcel of gas advected around the planet is slow compared to the condensation timescale. When a parcel of gas is transported from a hot to a cold part of the atmosphere, the most refractory condensates, such as $\rm MgSiO_3$ form first, depleting the gas in cloud forming elements such as $\rm Si$. When the temperature drops low enough, more volatile compounds such as $\rm SiO_2$ are unable to form since the surrounding gas is depleted in $\rm Si$~\citep[e.g.][]{Visscher2010}. If the opposite is assumed, i.e. if the parcels of gas are supposed to move faster than the growing of the grains, then all condensates form at the same time, leading to a prevalence of more volatile compounds such as $\rm SiO_2$~\citep[e.g.][]{Helling2008}.

Here we use the~\emph{equilibrium cloud} framework. In this approach, only a handful of condensates have high enough abundances and opacities to form optically thick clouds in solar composition atmospheres~\citep{Marley2000,Morley2012}. We consider the effect of $\rm CaTiO_{3}$ (perovskite), $\rm Al_2O_3$ (corundum), $\rm Fe$ (iron), $\rm MgSiO_3$ (enstatite), $\rm Cr$ (chromium), $\rm MnS$ (manganese sulfide) and $\rm Na_2S$ (sodium sulfide). Given that the optical properties and condensation curves of $\rm MgSiO_{3}$ and $\rm Mg_{2}SiO_{4}$ are very similar~\citep{Wakeford2015} we use $\rm MgSiO_{3}$ as a representation of both types of silicate clouds. Other condensates such as $\rm KCl$, $\rm ZnS$, $\rm H_2O$ form at lower temperatures and are never present in the dayside of our models. Therefore they cannot affect the Kepler lightcurve of the planet and are neglected in this study. When several cloud species are present, clouds of different compositions are considered independent. The scattering and absorption optical depths are summed and the asymmetry parameters are averaged using the scattering optical depth of each gas as a weighting function.}

We consider monosize particle clouds: all condensable cloud material condenses into particles of size $a$\footnote{We actually use a narrow log-normal distribution with a width of $\sigma=1.05$ around the mean particle size in order to smooth the Mie bumps in the opacities.}. We assume that the cloud is efficiently mixed between the cloud deck and the cloud top pressure $p_{\rm top}$. When a condensable species is present, its atomic abundances are constant and set by the initial condition, assumed solar. Inside a grid cell the number of atoms in condensed form is determined such that the the partial pressure of the remaining gas is equal to the saturation pressure. At pressures lower than the cloud top level $p_{\rm top}$, the atmosphere is set to be devoid of clouds. The cloud top level and the particle size are free parameters that parametrize the vertical mixing and the microphysics respectively. For our fiducial model we choose the values of $a=\SI{0.1}{\micro\meter}$ and $p_{\rm top}=\SI{1}{\micro\bar}$, leading to the strongest signature of clouds. Our main conclusions are independent of this choice (see Section~\ref{sec::size}).

\section{Comparison with the data: models with a single cloud species}
\subsection{Kepler phase curves as a probe of the cloud composition}
\subsubsection{Thermal structure and outgoing flux}
\label{sec::Fluxes}

\begin{figure*}[] 
 \centering
 \includegraphics[width=0.9\linewidth]{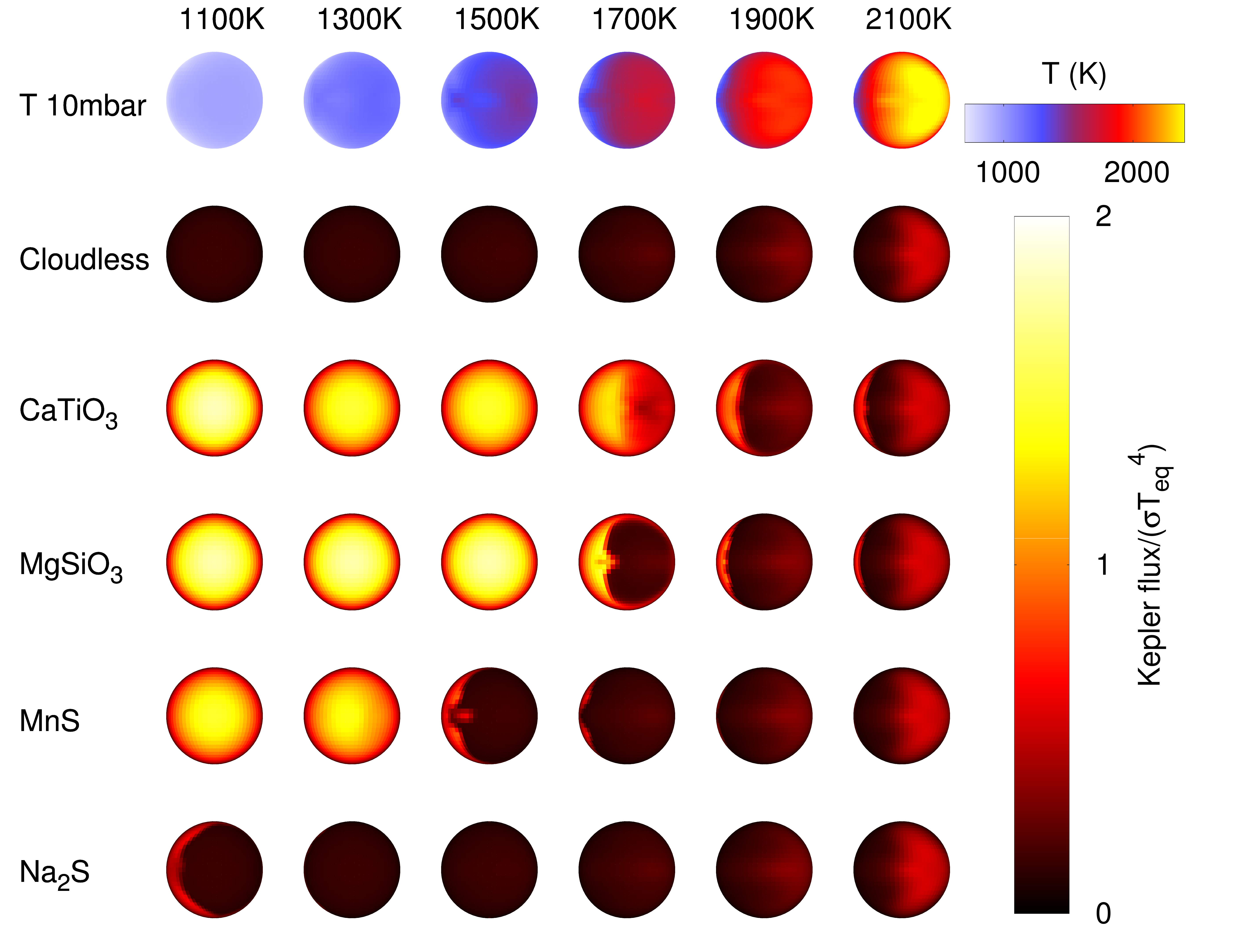}
 \caption{
 Temperature and outgoing flux from the dayside of hot Jupiters with different equilibrium temperatures. The first row shows the temperature at $\SI{10}{\milli\bar}$ as calculated by our global circulation model. The following rows show the total flux (emitted + reflected) from the dayside hemisphere of the planet in the spectral range observed by the \emph{Kepler} spacecraft in units of $\sigma T_{\rm eq}^4$. The second row is a model without clouds whereas in the subsequent rows one cloud species is condensing. We use $ P_{\rm top}=\SI{1}{\micro\bar}$ and $a=\SI{0.1}{\micro\meter}$.}
 \label{fig::maps}
\end{figure*}

We run the SPARC/MITgcm to calculate the three-dimensional thermal structure of Jupiter-size, solar-composition, tidally locked planets orbiting a solar-type star with an equilibrium temperature ranging {\ct from $1000\,\rm K$ to $2200\,\rm K$} with a step of $100\,\rm K$ and show the resulting temperature maps in the first row of Figure~\ref{fig::maps}. 

Two trends emerge from these temperature maps. First, the temperature variation across the dayside increases with the equilibrium temperature, as does the day/night temperature contrast~\citep{Perna2012,Perez-becker2013a,Komacek2016}. {\ct As a consequence, when the equilibrium temperature is multiplied by a factor of 2.2, the maximum temperature of the dayside at $\SI{10}{\milli\bar}$ increases by a factor 3 while the minimum temperature of the dayside increases by only a factor 1.5, with the western limb being the coolest part of the dayside.} The second important trend is the asymmetry in temperature in the dayside: the hottest hemisphere is not centered on the substellar point but is shifted eastward and this shift decreases when increasing the equilibrium temperature: the time needed for a parcel of gas to reach radiative equilibrium is inversely proportional to the temperature to the third power~\citep{Showman2002}, meaning that hotter planets have a dayside temperature distribution closer to the radiative equilibrium and thus more symmetrical.

The second row of Figure~\ref{fig::maps} shows the flux from the dayside of the planet in the \emph{Kepler} bandpass for cloudless planets. In that case scattering by the gas is small compared to the absorption by alkali atoms~\citep{Sudarsky2000} and the thermal contribution to the lightcurve is always dominant. The flux map therefore tracks the temperature map. At low equilibrium temperature, the planet is too cold to radiate in the \emph{Kepler} bandpass. As the equilibrium temperature increases, the thermal emission becomes significant because the gas emits more light and emits it at shorter wavelengths that overlap with the \emph{Kepler} bandpass.

The last four rows of Figure~\ref{fig::maps} show the flux from the dayside of the planet in the \emph{Kepler} bandpass assuming that different types of clouds are present in the atmosphere. In that case scattering is important at low temperature whereas thermal emission dominates at high temperatures. The appearance of the planet's dayside always follows the same trend: clouds cover the whole dayside at low equilibrium temperatures then disappear from the eastern part of the dayside when the temperature becomes too hot and finally are pushed toward the western limb where they remain even at large equilibrium temperatures. The equilibrium temperature at which the transition from a fully cloudy to a partially cloudy planet happens is a strong function of the condensation temperature of each species (see also Figure~\ref{fig::opd} in appendix).

\subsubsection{Cloud composition}
\label{sec::comp}
\begin{SCfigure*}[][t!]
 \centering
  \includegraphics[width=1.4\linewidth]{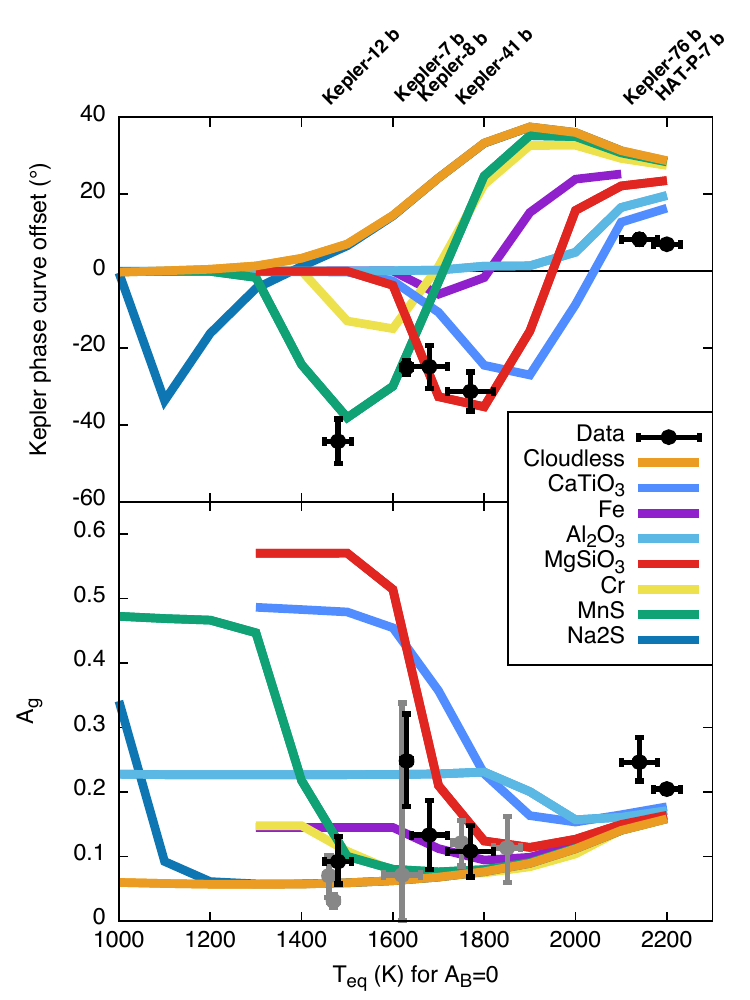}
\caption{
Offset of the maximum of the \emph{Kepler} lightcurve compared to the phase of the secondary eclipse (top) and apparent geometric albedo of the dayside hemisphere (bottom). Each line represents either a model without clouds or a model with a single cloud species. The cloud particle size is assumed to be $\SI{0.1}{\micro\meter}$ and the cloud top pressure is $\SI{1}{\micro\bar}$. The data is from~\cite{Esteves2015}, black points have measured shift whereas grey points only have an albedo measurement.}
 \label{fig::offsets}
\end{SCfigure*}

We use the flux maps of Figure~\ref{fig::maps} to model the planetary lightcurves in the Kepler bandpass. From these lightcurves we extract the offset of the maximum compared to the secondary eclipse and the apparent geometric albedo of the planet's dayside. Both quantities are plotted in Figure~\ref{fig::offsets} together with the \emph{Kepler} data from~\cite{Esteves2015}. 

For cloudless planets, the flux map follows the temperature map and, because the temperature map is always shifted eastward, the offset in the \emph{Kepler} lightcurve is always positive. The magnitude of the shift in the \emph{Kepler} lightcurve is determined by two competitive physical processes: the temperature contrast and the shift of the hottest point eastward of the substellar point. While the temperature contrast increases with equilibrium temperature, the eastward shift of the hot spot decreases. At these temperatures, the hotter the gas, the greater the emission in the \emph{Kepler} bandpass. Therefore an increase in the temperature contrast leads to an increase in the shift of the maximum of the \emph{Kepler} lightcurve. On the contrary, a smaller shift of the hottest point leads to a smaller shift of the maximum of the lightcurve. For equilibrium temperatures smaller than $1900\,\rm K$ the effect of the flux contrast dominates and the shift of the maximum of the Kepler lightcurve increases with equilibrium temperatures. For hotter planets, the effect of the shift of the hot spot dominates and the shift of the maximum of the \emph{Kepler} lightcurve decreases with equilibrium temperatures.

For cloudy planets, the offset in the \emph{Kepler} lightcurve is always zero at low equilibrium temperatures, when the planet is homogeneously covered by clouds. It reaches a minimum when inhomogeneous clouds are present and becomes positive at high temperatures when the thermal emission dominates the flux. The equilibrium temperature for which there is a shift depends weakly on the physical properties of the cloud, such as particle size and vertical extent (see Section~\ref{sec::size}), and is therefore a good signature of the cloud composition.

The bottom panel of Figure~\ref{fig::offsets} shows the apparent geometric albedo in the \emph{Kepler} bandpass. Here we use the word "apparent" as it is calculated directly from the planets' secondary eclipse depth: it is the ratio of the flux received by the planet from the star to the total outgoing flux leaving the planet in our direction. The apparent albedo is determined both by the reflectivity of the atmosphere and by the thermal emission of the planet in the \emph{Kepler} bandpass. For cloudless planets the apparent geometric albedo increases with equilibrium temperatures because more photons are emitted in the \emph{Kepler} bandpass when the planet gets hotter. For cloudy planets, the albedo follows the same curve as for the cloudless planet at high equilibrium temperatures, when the planet's dayside is too hot for clouds to form. When the equilibrium temperature decreases, the albedo increases as the cloud cover in the dayside increases. When the planet is cold enough to be homogeneously covered by clouds, the apparent geometric albedo reaches a plateau, the value of this plateau being determined by the abundance and the scattering properties of the condensates. 

By comparing the maps of Figure~\ref{fig::maps} and the two panels of Figure~\ref{fig::offsets}, we see that large phase shifts in the \emph{Kepler} phase curves appear for planets that only possess a longitudinally narrow layer of cloud near the western terminator (see the case of MnS clouds with $T_{\rm eq}=1500\,\rm K$) or for planets that have half their dayside covered by clouds (see the case of silicates clouds with $T_{\rm eq}=1700\,\rm K$). This corresponds to planets with low to moderate albedos. Planets with large geometric albedos must be entirely covered by clouds and thus we do not expect them to have a significant shift in their phasecurve.

{\ct Few cloud species have a single scattering albedo large enought to produce a large shift in the \emph{Kepler} lightcurve.  {\ct Perovskite clouds,} silicate clouds, manganese sulfide clouds or sodium sulfide clouds have a single scattering albedo larger than 0.95 and can produce a large offset in the \emph{Kepler} lightcurve. Corundum clouds ($w_0\approx0.88$), iron clouds ($w_0\approx0.67$) and chromium clouds ($w_0\approx0.67$) have smaller albedos, leading to darker clouds and a smaller phase shift in the Kepler bandpass. 

The integrated vertical optical depth of our cloud species (see Figure~\ref{fig::opd} in appendix) are several orders of magnitudes larger than shown in the brown dwarf literature. This is expected as the cloud optical depth at a given pressure level is inversely proportional to the gravity of the object and hot Jupiters have gravities $\approx100$ times smaller than brown dwarfs~\citep[see][]{Marley2000}. As a consequence, clouds that are believed to be optically thin or barely optically thick in brown dwarfs atmospheres, such as perovskite clouds, chromium clouds or sulfide clouds~\citep[see][]{Morley2012} can play an important role in hot Jupiters atmospheres.}

 As shown in Figure~\ref{fig::offsets}, models with silicate or manganese sulfide clouds can match the albedo and phase shift of Kepler-7b and Kepler-8b. Models with silicate clouds are also able to reproduce the observations of Kepler-41b but predict a zero phase shift and a large albedo for Kepler-12b, in stark contrast with the observations. Conversely, models with manganese sulfide clouds match the observations of Kepler-12b but are unable to reproduce the large phase shift observed for Kepler-41b as the planet is too hot for MnS clouds to form in its dayside. Models with {\ct perovskite clouds can also reproduce the phase shift of Kepler-41b, but predict a larger than observed albedo.}

For the hotter planets Kepler-76b and HAT-P-7b, models with {\ct perovskite clouds or corundum clouds} provide the best match to the observations, although the models predict higher positive phase shifts and lower apparent geometric albedos than observed. 

{\ct Physical mechanisms not taken into account in our fiducial set of simulations can explain this discrepancy. The presence of titanium dioxide in the atmosphere of high equilibrium temperature can enhanced the dayside temperatures and lowering the thermal inertia of the atmosphere, leading to higher apparent albedos and smaller shifts in the lightcurves (see Sec.~\ref{Sec::TiO}). The presence of magnetic drag for high equilibrium temperature planets~\citep[e.g.][]{Perna2010,Rauscher2013,Batygin2013,Rogers2014,Rogers2014a} could reduce the speed of the eastward jet, leading to a decrease of the observed shift and an increase of the apparent albedo.}

We do not find a model with a single cloud species that is able to match \emph{all} current observations. Models with manganese sulfide clouds provide our best match for the coolest planets but cannot reproduce the observations of hotter planets. Conversely, silicate clouds seem necessary to match the observations of the planets with intermediate equilibrium temperatures but fail when they are used to interpret the data from the coolest planet. Finally, both cloud species seem unable to reproduce the observations of the hottest planets,  while {\ct perovskite} or corundum clouds provide a solution closer to the observations.

\subsection{Model sensitivity to parameters and assumptions}
\subsubsection{Particle size and cloud top level}
\label{sec::size}
\begin{figure*}[] 
 \centering
  \includegraphics[width=\linewidth]{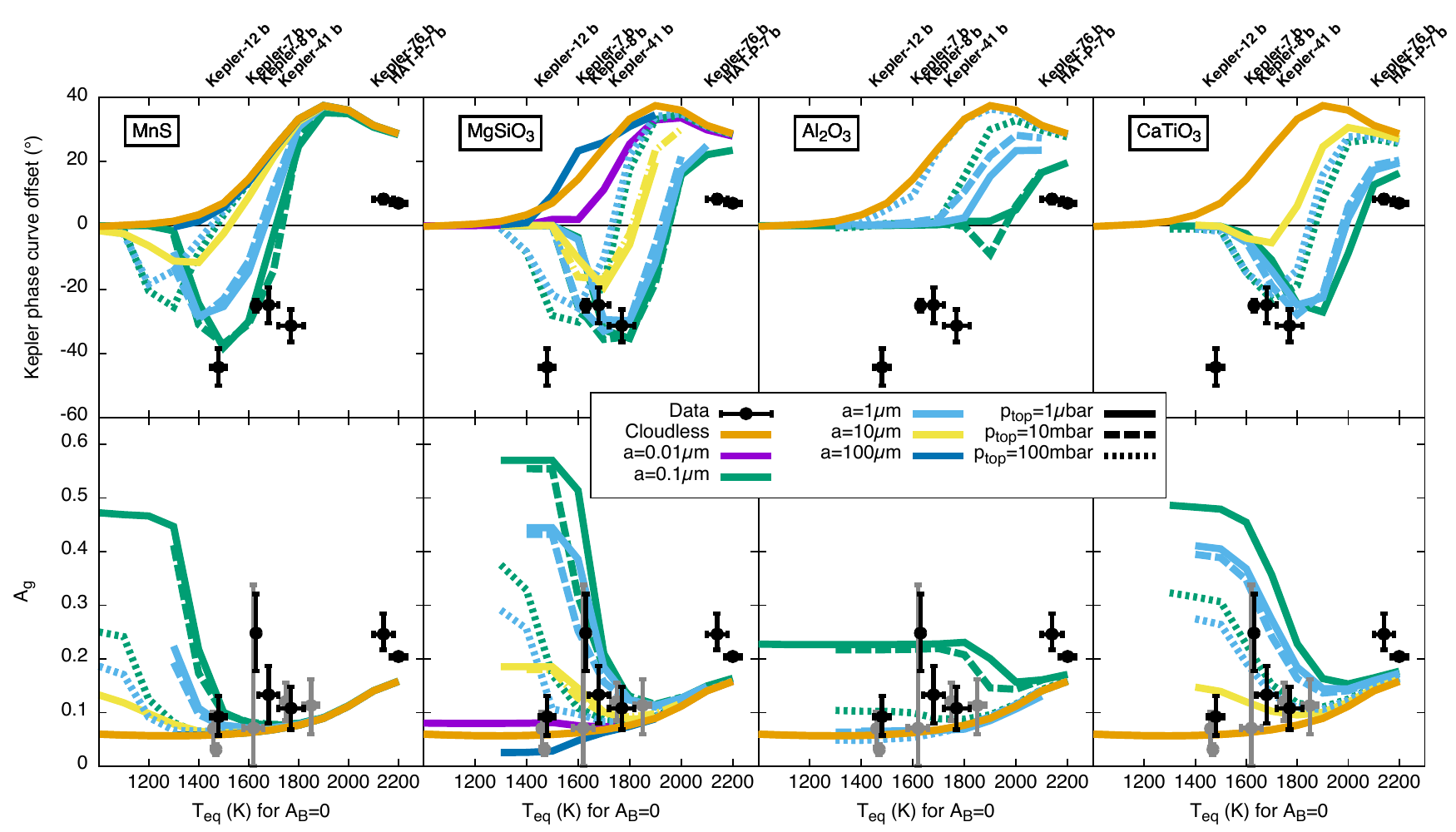}
\caption{
{\ct Phase shift of the maximum of the \emph{Kepler} lightcurve relative to the secondary eclipse for cloudless and cloudy models with different cloud particle size and different cloud top pressure assuming equilibrium clouds made of $\rm MnS$ (left panel), $\rm MgSiO_{3}$ (middle left panel), $\rm Al_2O_3$ (middle right panel) or $\rm CaTiO_3$ (right panel).}}
 \label{fig::ptop-a}
\end{figure*}

The particle size and vertical distribution of clouds in hot Jupiters is poorly constrained. From the slope of the optical transmission spectrum, the maximum particle size at the limb appears to be sub-micron for several planets~\citep{LecavelierDesEtangs2008,Sing2013}, yet it is unclear how this constraint can translate to the dayside at the higher pressures probed by the phasecurve method. From complex microphysical models the cloud particle size and vertical distribution can be calculated a priori at any location of the planet. These microphysical models, however, appear very sensitive to the temperature and vertical mixing~\citep{Lee2015}, making the predictions for a given planet difficult to generalize when studying planets over a range of equilibrium temperatures. Here we explore the sensitivity of our conclusions to the particle size and vertical mixing for the most important cloud species in our model $\rm MnS$, $\rm MgSiO_{3}$, $\rm Al_2O_3$ and $\rm CaTiO_3$.

We vary the particle size from $\SI{0.01}{\micro\meter} $ to $\SI{100}{\micro\meter}$. As seen in Figure~\ref{fig::ptop-a}, the phase curves for $\SI{0.01}{\micro\meter}$ and $\SI{100}{\micro\meter}$ are similar to the cloudless phase curves. For small particle sizes, in the Rayleigh regime, particle opacities are proportional to the radius of the particle to the power of 6 whereas their total number inside the cloud is proportional to the particle size to the power -3. Therefore we expect the clouds to become transparent at small particle sizes. For the large particle sizes, in the geometric optics limit, particle opacities become proportional to the radius of the particle to the power of 2 and the particles become less reflective due to the increased importance of forward scattering~\citep{Cuzzi2014}. As a consequence, clouds become poor reflectors at large particle sizes (note that absorption can remain important). We find that an optimal particle size of $\SI{0.1}{\micro\meter}$ produces the largest shift in the \emph{Kepler} lightcurve and the largest apparent albedo in the \emph{Kepler} bandpass. Changing the particle size affects the magnitude of the shift of the lightcurve but does not affect the equilibrium temperature for which this shift happens since this is given directly by the thermal structure and the condensation curve of each species.

Although theoretical work has tried to understand the mixing processes in the radiative atmosphere of hot Jupiters~\citep{Parmentier2013}, no observations have yet been able to constrain the mixing rates in these planets. Whereas the pressure of the cloud base is determined by the condensation curve, the vertical extent of the cloud and thus the cloud top pressure is determined by the strength of the vertical mixing. Here we vary the cloud top pressure from $\SI{1}{\micro\bar}$ to $\SI{100}{\milli\bar}$.  As seen in Figure~\ref{fig::ptop-a} there is almost no sensitivity to the cloud top pressure between $P_{\rm top}=\SI{1}{\micro\bar}$ and $P_{\rm top}=\SI{10}{\milli\bar}$: the clouds above the 10mbar level are not abundant enough to become optically thick and the photosphere is unchanged when varying $P_{\rm top}$. When $P_{\rm top}$ reaches $\SI{100}{\milli\bar}$, the photosphere is at a deeper level, where the atmosphere is hotter. The cloud distribution at the photosphere is therefore changed and the offset curve shifts to lower equilibrium temperatures. For larger $P_{\rm top}$ the cloud top is below the photosphere and the phase curve is similar to the cloudless case.

As a conclusion, silicate, manganese sulfide and perovskite clouds produce a significant shift in the Kepler phase curve for particle sizes of $\SI{0.1}{\micro\meter}$ to $\SI{10}{\micro\meter}$ and cloud top pressure from $\SI{1}{\micro\bar}$ to $\SI{100}{\milli\bar}$. Corundum clouds only affect the observations for particle sizes of $\SI{0.1}{\micro\meter}$ and are transparent for other particle sizes. Finally, by varying the particle size by four orders of magnitude and varying the cloud top pressure by six orders of magnitude there is no model with a single cloud species that can match all current \emph{Kepler} phasecurve shifts and albedos. 

\subsubsection{The case of  gaseous TiO}

\label{Sec::TiO}

If present in solar abundance in irradiated atmospheres, titanium and vanadium oxides should produce a strong thermal inversion by absorbing a significant amount of the stellar light in the upper atmosphere and by impeding the thermal cooling of the atmosphere~\citep{Hubeny2003,Fortney2008,Showman2009,Parmentier2015}. To date, the signature of TiO has been found in the two of the hottest exoplanets~\citep{Haynes2015,Evans2016} but no convincing evidence has been found for cooler ones~\citep{Desert2008,Huitson2013}. Given the strong change in the radiative forcing expected from this molecule, we calculated a set of global circulation models incorporating TiO in solar composition. 

{For the cloudless case, at low equilibrium temperatures, the offset of the \emph{Kepler} phase curve is negative rather than positive as in the case without TiO/VO. At these low equilibrium temperatures, the lightcurve is dominated by Rayleigh scattering from the gas. On the east of the dayside, where the temperatures are hotter due to the eastward shift of the hot spot the presence of TiO/VO increases the absorption of the stellar light and reduces the reflectivity of the atmosphere, leading to a negative shift in the lightcurve. At large equilibrium temperature, the offset of the \emph{Kepler} phase curve is smaller than in the case without TiO/VO. In the models with TiO/VO, the stellar flux is absorbed at lower pressures, where the radiative timescale is small. As a consequence the eastward shift of the hot spot is smaller in the case with TiO/VO~ \citep{Fortney2008,Showman2009}, leading to a smaller shift in the Kepler lightcurve.

For the cloudy cases, the shifts are smaller than in the case without TiO and are present at lower temperatures: the presence of TiO maintains a higher temperature contrast on the dayside but a smaller eastward shift of the temperatures. As a result, asymmetric cloud distributions are more common, but this asymmetry is smaller. Interestingly, the shifts produced by silicate and perovskite clouds are very similar, although the condensation curve of the two materials are differ by $\approx200\,\rm K$. Models with TiO have a sharper temperature contrast, meaning that clouds with different condensation temperatures can have a similar horizontal distribution and thus produce a similar phase shift in the light curve. 

{\ct Whereas the presence of perovskite clouds is a possible explanation for the high shift in the Kepler lightcurve of Kepler-41b when gaseous TiO is not used to calculate the atmospheric opacities, it is no more a good explanation when the opacity of gaseous gaseous TiO is used in the calculation. If perovskite clouds were present in enough quantity to produce an important shift in the Kepler lightcurve, it should also be present in enough quantity for gaseous TiO to be an important absorber in the cloudless part of the atmosphere. As a consequence, perovskite clouds seem a less likely candidate to explain the offset of Kepler-41b than silicate clouds.}
 
Overall, the models with gaseous TiO can possibly explain the low positive shifts and large apparent albedos observed for the two hot planets Kepler-76b and HAT-P-7b without invoking the presence of clouds. For the cooler planets, however, the models with TiO provide a worse match to the data, pointing toward a lack of TiO in cool planets, in agreement with current observations. The cold trapping of TiO in the deep layers planets with an equilibrium temperatures lower than $\approx 1900\,\rm K$, when the deep temperature contrast crosses the condensation curve of TiO as predicted by our models (see Section~\ref{sec::CT} and Appendix), provides a natural explanation for the lack of observed signatures in cool planets~\citep[see also][]{Spiegel2009}.

\begin{figure}[htb] 
 \centering
  \includegraphics[width=\linewidth]{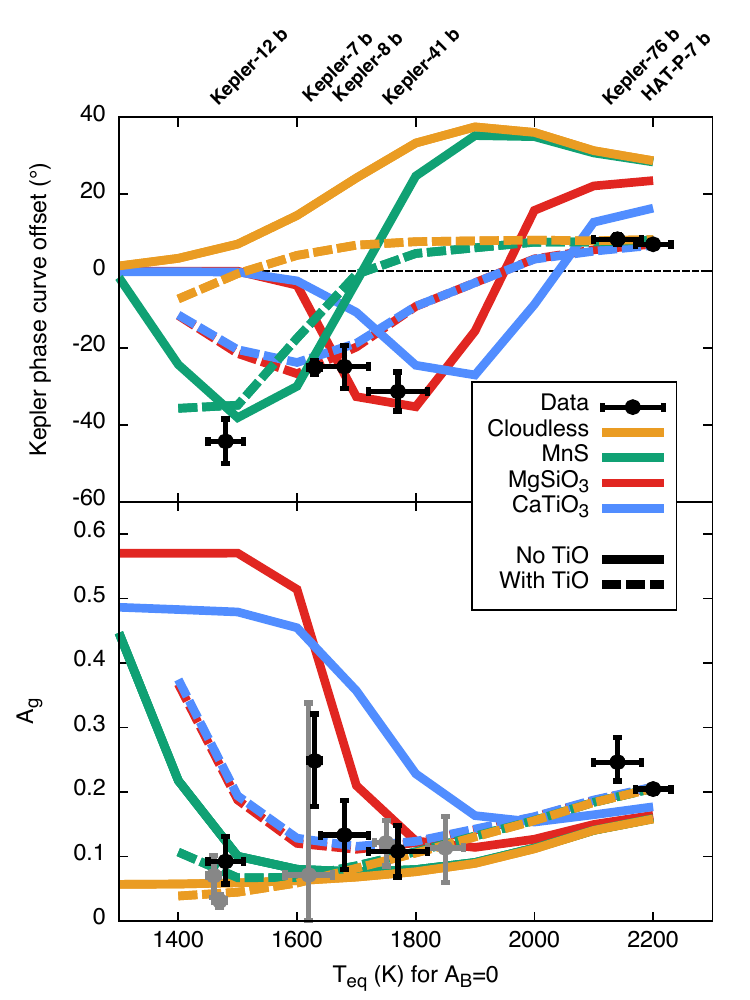}
\caption{
Phase shift of the maximum of the \emph{Kepler} lightcurve relative to the secondary eclipse (top) and apparent geometric albedo of the planet in the \emph{Kepler} bandpass (bottom) for cloudless and cloudy planets with enstatite clouds. Plain lines models are our fiducial models, without TiO. Dashed line are models where TiO has been added in equilibrium abundances both in the GCM and in the post-processing calculations.}
 \label{fig::TiO}
\end{figure}

\subsubsection{Cloud feedback}
\label{sec::cloudsFB}
Clouds can significantly affect atmospheric opacities and thus the atmospheric circulation and the thermal structure of irradiated planets. Our fiducial models, however, do not consider this feedback: we use a cloudless global circulation model to determine the distribution of clouds on the planet and calculate the lightcurve. This allows us to estimate the shift in the \emph{Kepler} phase curve for a large number of cloud species, particle sizes and cloud top pressures without the cost of running a computationally demanding global circulation model for each cloud scenario. 

In order to estimate the importance of the cloud feedback on the circulation, we calculated a new set of global circulation models including the same cloud setup as described in Section~\ref{Sec::Clouds}. The spatial distribution of clouds is determined by comparing the condensation curve of a given species and the thermal structure of the planet. Both absorption and scattering by the clouds are taken into account by the GCM, leading to a self-consistent, cloudy, global circulation model. For simplicity we do not use trace particles advected by the flow as in~\cite{Charnay2015a}  to represent the mixing of the cloud particles by the circulation. Instead we use a similar approximation as in the previous section where the cloud abundance is determined based on the local pressure and temperature : all condensable material condenses into particles of size $a$ so that the remaining partial pressure of the condensable gas matches its saturation pressure. At pressures lower than $P_{\rm top}$, the atmosphere is set to be devoid of clouds.

We consider the case of $\rm MnS$ clouds assuming a cloud top pressure of $\SI{1}{\micro\bar}$ and a particle size of $\SI{0.1}{\micro\meter}$, leading to the strongest possible forcing due to $\rm MnS$ clouds. Clouds have two main effects: they increase the scattering at short wavelengths, increasing the albedo of the planet leading to an overall reduction of the temperatures. They also increase the scattering and absorption in the thermal wavelengths, leading to a larger greenhouse effect and thus to higher temperatures. When the dayside has a mainly clear sky but the nightside is covered by clouds the greenhouse effect is dominant. The same amount of radiation penetrates into the atmosphere on the dayside whereas the radiative cooling on the nightside is reduced. As a consequence, the temperature of the whole atmosphere increases, leading to a hotter dayside. At a given equilibrium temperature, the cloud coverage of the dayside is smaller in a model with cloud feedback than in a model without it. As seen in Figure~\ref{fig::shift-SC}, the result is a displacement of the offset and albedo vs. equilibrium temperature curves by $\approx100\,\rm K$ towards lower equilibrium temperatures. {\ct For the planets that have a mostly clear dayside and a cloudy nightside, for which an offset in the Kepler lightcurve can be measured, we show that the cloud feedback is of second order on the circulation compared to the effect of the irradiation, leading to a small change in the estimated phase offset and albedo of the planet. When the planet is fully covered by clouds, the effect on the atmospheric circulation can be larger (see e.g. ~\citet{Dobbs-dixon2013} and \citet{Charnay2015a}). However, when clouds are homogeneously covering the planet dayside, the phase offset must be zero and the albedo rather independent on the circulation. As a consequence, our conclusions based on the cloudless temperature map remain valid.}

\begin{figure}[t!] 
 \centering
  \includegraphics[width=\linewidth]{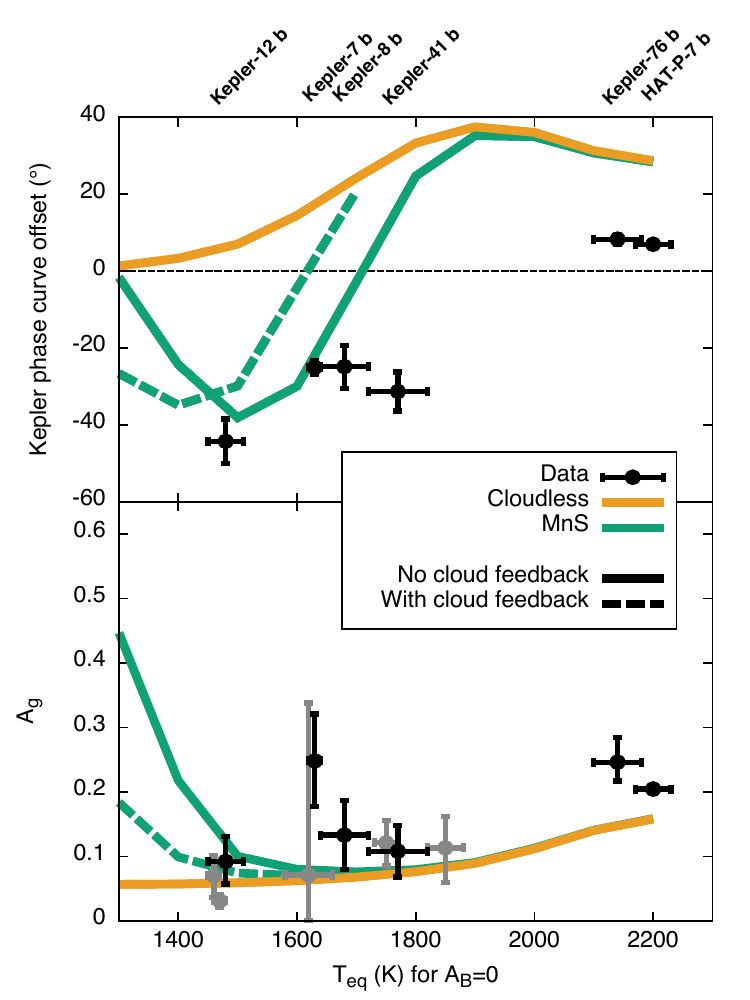}
\caption{
Phase shift of the maximum of the \emph{Kepler} lightcurve relative to the secondary eclipse for self consistent global circulation models (red line) and for models where the cloud feedback on the circulation is not taken into account (green line).}
 \label{fig::shift-SC}
\end{figure}

\subsubsection{Metallicity}
Our previous models assumed a solar-composition for the planet's atmospheres. However, planets are expected to be enriched in metals compared to their parent-star. In our solar-system, the metal enrichment is correlated with the planetary mass, the lightest planets being the most enriched ones. This trend seems to hold for exoplanets, with a measured metallicity of $0.4-3.5\times$ solar for WASP-43b~\citep{Kreidberg2014}. The planets in our sample have masses ranging from $\sim0.5$ to $\sim2\,\rm M_{\rm J}$. Based on the trend described in~\citep{Kreidberg2014}, a metal enrichment between $\sim1-5\times$ solar seems plausible. 

The metal content of the atmosphere should affect the \emph{Kepler} lightcurve in several ways. A metal enriched atmosphere contains more cloud material, enhancing the probability of having thick clouds. It also contains a larger abundance of gaseous absorbers such as sodium or potassium, leading to a darker atmosphere particularly on the cloudless spots. 

The metal content of the atmosphere also affects the thermal structure of the planet: in a metal-rich atmosphere the stellar light is absorbed at lower pressures, leading to a hotter upper atmosphere and a cooler deep atmosphere. As shown in~\cite{Kataria2015}, a metal enrichment of $5\times$ solar should increase the temperature on the dayside by $\approx100\,\rm K$ compared to a solar composition. This should translate to a shift in the offset and albedo vs. equilibrium temperature curves of Figure~\ref{fig::offsets} towards cooler equilibrium temperatures by $\approx100\,\rm K$. 

{\ct The layer where the energy is deposited has a shorter radiative timescale in the metal-rich case, because it has a lower pressure and a higher temperature than in the solar composition case. Therefore, a smaller eastward shift of the hot spot is expected, leading to smaller westward shift of the cloudiest hemisphere and a smaller shift of the~\emph{Kepler} lightcurve. Because the radiative timescale is proportional to the temperature to the power of three, the change in the shift of the hot spot should be larger for the hotter planets.}

Another effect of metal enrichment is to shift the condensation curves of condensable species to higher temperatures\footnote{The condensation curve is the pressure-temperature line for which the partial pressure of the gas is equal to its saturation pressure. Increasing the metallicity of the atmosphere increases the partial pressure of the condensing gas and shifts the condensation curves to higher temperature.}. As shown in~\cite{Kataria2015}, a metal enrichment of  $5\times$ solar shifts the condensation curve of condensable species by$\approx100\,\rm K$ towards higher temperatures, which should shift the offset and albedo curves of Figure~\ref{fig::offsets} to higher equilibrium temperatures by $\approx100\,\rm K$. 


{\ct In order to understand the trade-off between these different processes, we ran a set of global circulation model with a five time solar metallicity. We then calculate the lightcurves assuming the same metal enrichment to determine the opacities, the abundances of cloud-forming material and the condensation curves for $\rm MnS$ and $\rm MgSiO_{3}$ clouds~\citep{Morley2012,Visscher2010}. As shown in Figure~\ref{fig::shift-SC}, for the cloudless case, the metal-rich atmosphere has a larger shift and a smaller albedo when the equilibrium temperature is cooler than $\approx1800\,\rm K$. The larger temperatures increase the flux emitted in the Kepler bandpass, leading to a larger flux contrast in the planet and thus a larger shift. This enhanced thermal flux is not large enough to compensate for the reduced Rayleigh scattering due to a larger number of optical absorbers and the planet appears darker than in the solar composition case. For planets hotter than $T_{\rm eq}\approx1800\,\rm K$, the enhanced thermal emission increases the apparent albedo whereas the reduced shift of the hot spot reduces the shift of the lightcurve. 

For cloudy planets, the enhanced metallicity do not change the equilibrium temperatures for which a shift is observed. This is because in a metal-rich atmosphere both the temperature and the condensation curves are shifted by $\approx100\,\rm K$. As a consequence the cloud distribution remain unchanged. When silicate clouds are present, the lightcurve shifts are predicted to be smaller than in the solar-composition case. This is due to the smaller eastward shift of the hot spot. This does not happen for the manganese sulfide clouds since for cooler planets, the change in the shift of the hot spot is smaller than in hotter planets. The higher abundance of cloud material and the higher abundance of absorbing gas phase molecules compensate so that the apparent albedos in the \emph{Kepler} bandpass remain unaffected. 

As a conclusion, we find that the metallicity of the atmosphere does not affect significantly the shift of the lightcurve and the the planet apparent albedo. In particular, an enhanced metallicity does not allow us to find a model with a single cloud species explaining all current observations.
}

\begin{figure}[t!] 
 \centering
  \includegraphics[width=\linewidth]{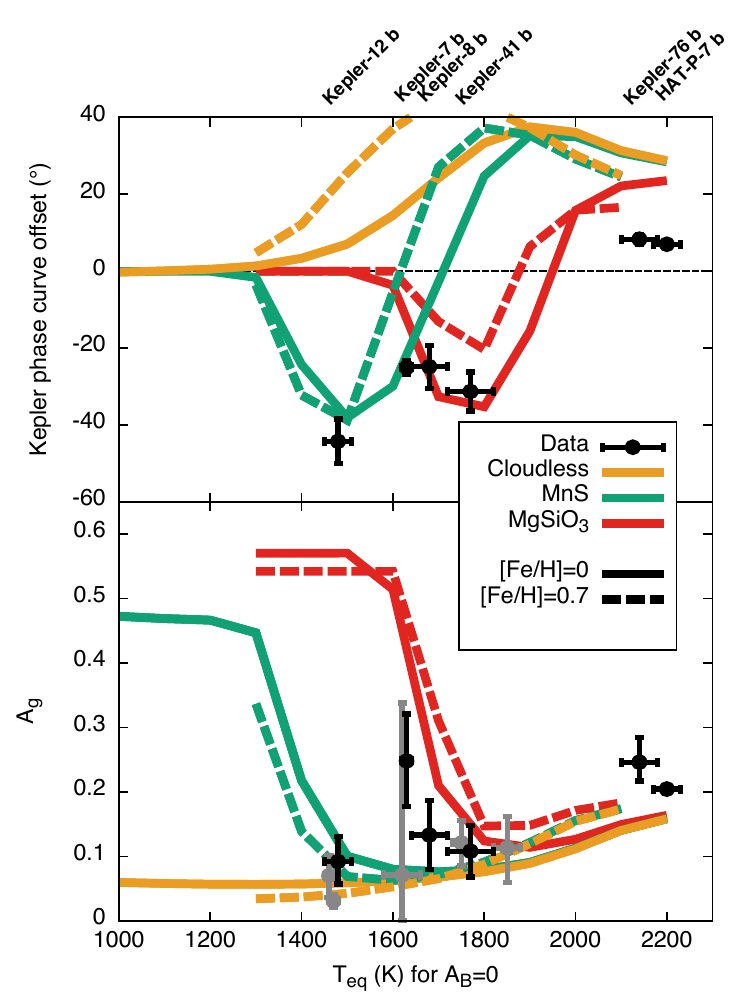}
\caption{
Phase shift of the maximum of the \emph{Kepler} lightcurve relative to the secondary eclipse for models with solar metallicity (plain lines) and models with a metallicity five times solar (dashed lines).}
 \label{fig::shift-SC}
\end{figure}

\section{Models with several cloud species}
\label{sec::CT}
\subsection{Disappearance of silicate clouds: a possible L/T-like transition}

{\ct We showed in section~\ref{sec::comp} that different cloud composition are necessary to explain the offsets seen in different planets. Particularly, we showed that silicate clouds were favored to explain the observations of Kepler-41b whereas manganese sulfide clouds were necessary to reproduce the small albedo and large offset of Kepler-12b. We now wonder whether the presence of both clouds in all planets could explain the data. We calculate additional models where we assume that both cloud species form independent clouds and sum the contribution of each cloud to the extinction and scattering opacities. When both clouds are present, the offset and the albedo of the planet follows closely the one of the silicate clouds alone (red dashed line of Figure~\ref{fig::MnSMg}). Specifically, for equilibrium temperatures lower than $1600\,\rm K$ the model predicts a large albedo and zero shift, in contrast with the observed large phase shift of Kepler-12b and the low albedo of the three planets observed around an equilibrium temperature of $\approx1500\,\rm K$. 

We now investigate a scenario where silicate clouds rain out from the observable atmosphere when the deep temperature profile of the planet (at $10$ to $\SI{100}{\bar}$), as predicted by the GCM, crosses its condensation curve (dotted lines of Figure~\ref{fig::PT}, see also in Appendix). Although the upper atmosphere of hot Jupiters is believed to be well mixed by high velocity winds, the atmosphere is much quieter below the photosphere~\citep{Parmentier2013}. When the cloud base lies at pressures of ten to hundreds of bars, the weak vertical mixing in the $10-\SI{100}{\bar}$ region prevents the cloud material from reaching the photosphere, depleting the observable atmosphere both in the gaseous and the condensed phase of the cloud material. Given this assumption, silicate clouds are supposed to disappear from the observable atmosphere when the equilibrium temperature becomes cooler than $\approx1600-1700\,\rm K$.  As seen by the plain blue line of Figure~\ref{fig::MnSMg}, a model considering the presence of such a deep cold trap is able to reproduce current observations in the  $1400-1900\,\rm K$ equilibrium temperature range. Manganese sulfide clouds dominate the cloud composition for $T_{\rm eq}<1600\,\rm K$  leading to large shifts and low albedos whereas silicate clouds dominate the cloud composition for hotter planets.

The current dataset therefore suggests that silicate clouds are present in planets with an equilibrium temperature larger than $1600\,\rm K$ but that they are not present in cooler planets. Evidence for this transition is given by the large shift observed for Kepler-12b, the low albedo observed for the three planets around $T_{\rm eq}\approx1500\,\rm K$ and the lack of high albedo planets in the $1400\,\rm K-1600\,\rm K$ equilibrium temperature range. This transition is similar to the L/T transition in brown dwarfs~\citep{Kirkpatrick2005} but happens at slightly higher equilibrium temperatures ($\approx1600\,\rm K$ vs. $\approx1400\,\rm K$)\footnote{This can be explained by the presence of a large, quasi-isothermal radiative zone in the $\approx 1-\SI{100}{\bar}$ region of hot Jupiters. A convective brown dwarf and a hot Jupiter of the same equilibrium temperature should have similar temperatures where the optical depth reaches unity. At higher pressure, however, the temperature in the convective zone of a brown dwarf increases much faster than in the radiative zone of a hot Jupiter. Hot Jupiters are therefore cooler than brown dwarfs below the photosphere, making them more efficient at cold trapping cloud species.}.}

\begin{SCfigure*}[][htb]
 \centering
  \includegraphics[width=1.5\linewidth]{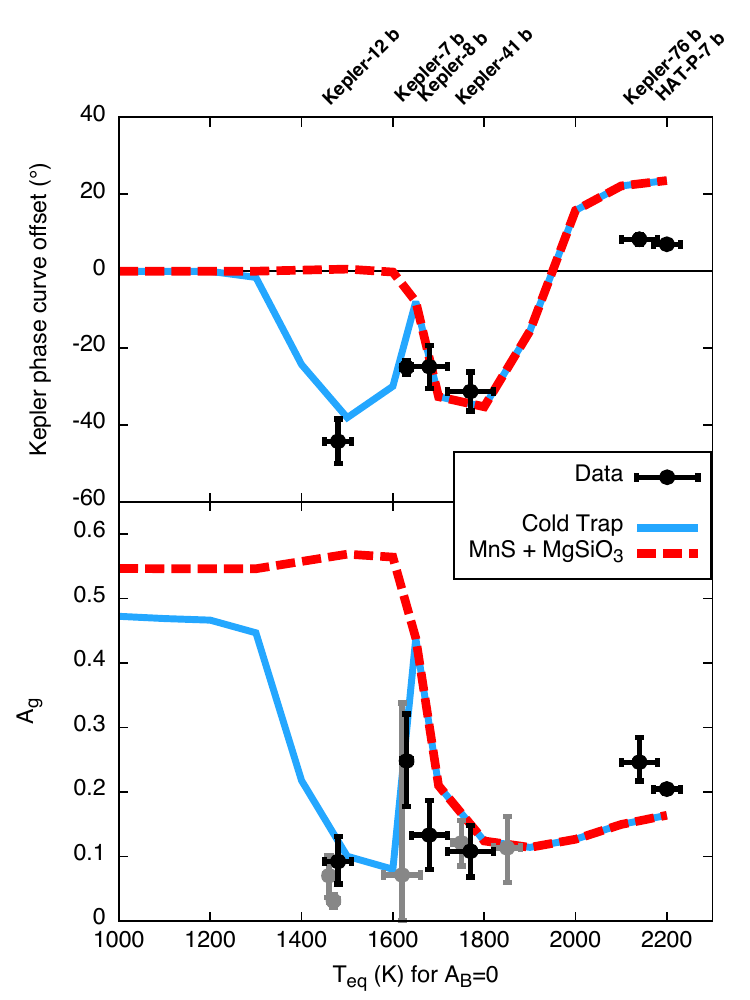}
\caption{
 Offset of the maximum of the \emph{Kepler} lightcurve (top) and apparent geometric albedo (bottom) as observed by~\cite{Esteves2015}. The dashed red line is a model where both MnS and $\rm MgSiO_3$ clouds are considered at the same time. The plain blue line is a model where the silicate clouds rain out from the atmosphere in planets cooler than $T_{\rm eq}=1600\,\rm K$. The particle size is $\SI{0.1}{\micro\meter}$ and the cloud top pressure is $\SI{1}{\micro\bar}$.}
 \label{fig::MnSMg}
\end{SCfigure*}

\subsection{Presence of a deep cold trap for all cloud species}

We now calculate additional models where the seven cloud species studied in this paper are considered at the same time. First, we consider the case where all cloud species are present at all equilibrium temperatures. As shown by the dashed lines of Figure~\ref{fig::CT}, in that case no significant shift in the lightcurve should be observed. The reflected flux is dominated by the most refractory clouds and adding other cloud species does not affect the overall flux. The most refractory clouds need to disappear from low-temperature planets in order to produce an observable phase shift.

\begin{SCfigure*}[][htb]
 \centering
  \includegraphics[width=1.5\linewidth]{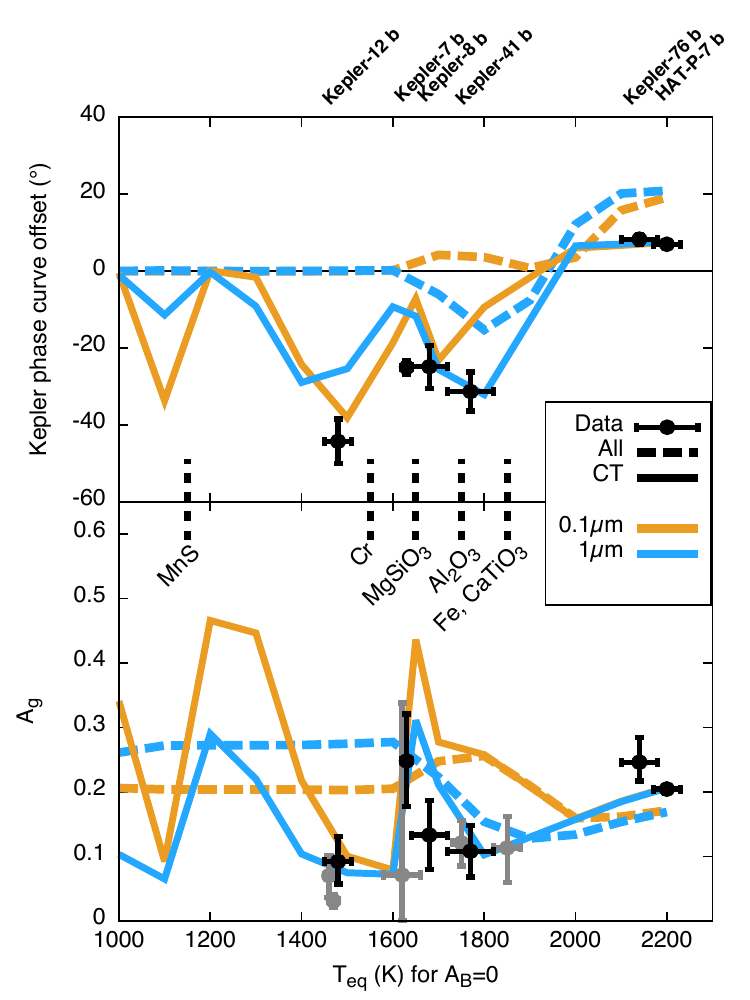}
\caption{
 Offset of the maximum of the \emph{Kepler} lightcurve (top) and apparent geometric albedo (bottom) as observed by~\cite{Esteves2015}. The dashed lines correspond to models where all cloud species are present at the same time whereas the plain lines correspond to models where the cloud and gaseous species present are determined using the cold trap model. The different colors are models assuming that the clouds are composed of different particle sizes. The cloud top pressure is assumed to be $\SI{1}{\micro\bar}$. The black vertical short-dashed lines show the approximate temperatures where various condensates are predicted to disappear from the observable atmosphere due to the deep cold trap at ~10-100 bars.}
 \label{fig::CT}
\end{SCfigure*}

{\ct We investigate a cold trap scenario where cloud species rainout from the observable atmosphere when the deep temperature profile of the planet (at $10$ to $\SI{100}{\bar}$), as predicted by the GCM, crosses its condensation curve (dotted lines of Figure~\ref{fig::PT}, see also the Appendix). In such a cold trap model, TiO is present in gas phase for $T_{\rm eq}>1900\,\rm K$ and rains out of the atmosphere for lower equilibrium temperatures. We predict the disappearance of iron and perovskite clouds between $1800\,\rm K$ and $1900\,\rm K$, of corundum clouds between $1700\, \rm K$ and $1800\,\rm K$, of silicate clouds between $1600\,\rm K$ and $1700\,\rm K$, of chromium clouds between $1500\,\rm K$ and $1600\,\rm K$ and of manganese sulfide clouds between $1100\, \rm K$ $1200\, \rm K$ (see the vertical dotted lines of Figure~\ref{fig::CT}). When particles sizes of $\SI{0.1}{\micro\meter}$ are assumed, the cold trap model reproduces well the offset and albedos of most observed planets. Particularly, the presence of gaseous TiO in the hottest models allows a better fit of the positive offset and large apparent albedo of Kepler-76b and HAT-P-7b. However, when $\SI{0.1}{\micro\meter}$ particles are considered, the model underestimates the offset and overestimates the albedo of Kepler-41b due to the presence of corundum clouds. 

When a larger particle size is assumed, the effect of corundum clouds is damped (see Sec.~\ref{sec::size}), leading to a better fit of the data over the $1700-1900\,\rm K$ temperature range. As a drawback, the effect of manganese sulfide clouds is also reduced, leading to a smaller than observed shift for Kepler-12b. As hotter planets have probably larger mixing rates than cooler ones their clouds might be composed of larger particles, providing a possible explanation for the remaining discrepancy between our models and the observations.

Several transitions in the cloud composition are predicted by our cold trap model. The exact equilibrium temperature at which these transitions happen are determined based on the deep temperature calculated assuming a solar-composition atmosphere in radiative equilibrium. However, given the large population of inflated hot Jupiters we know that the deep temperature profile is not solely determined by radiative equilibrium. Other physical processes, held responsible for the inflated nature of these planets, affect the deep temperature profile either by depositing heat directly in the 10-100 bar region (e.g. ohmic dissipation, mechanical greenhouse effect) and/or by decreasing the depth of the radiative/convective boundary (e.g. tidal dissipation)~\citep[see][for an overview]{Lopez2016}. As an example, ohmic dissipation is thought to increase the deep temperature profile by a hundred to a thousand kelvin depending on the strength of the planetary magnetic field~\citep{Perna2010,Spiegel2013}. The variation in the strength of these processes from planet to planet should impact the particular pressure-level of any deep cold trap, leading to a more blurry transition than shown by the models of this paper.

In conclusion, the current set of data cannot be reproduced if all cloud species are present in all planets. The cloud composition of hot Jupiter must depend on the equilibrium temperature of the planet. The presence of a cold trap in the $10-\SI{100}{\bar}$ region is a simple, reasonable and physically motivated explanation that allows us to reproduce all current observations. Because the presence of this cold trap for a given cloud species depends on the exact deep temperature profile, the determination of the cloud composition of a given planet could provide insights on the deep, usually unobservable, thermal structure of the planet.}


\section{Consequences for transit and secondary eclipse observations}

\begin{figure*}[t!]
 \centering
  \includegraphics[width=\linewidth]{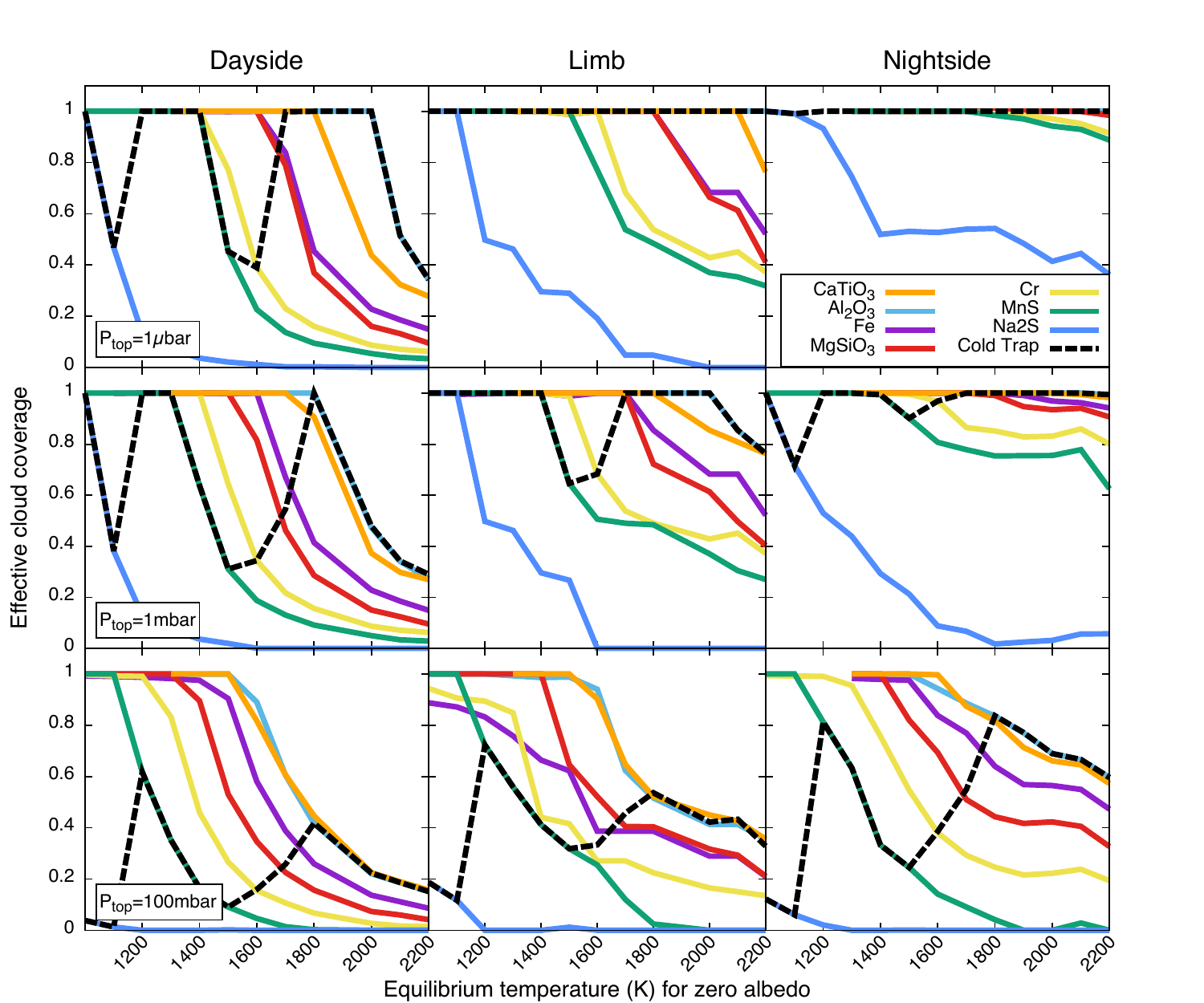}
\caption{Effective cloud coverage on the dayside (left column), at the terminator (middle column) and on the nightside (right column) of our modeled solar-composition tidally locked planets. The effective cloud coverage is the fraction of the planet covered by clouds optically thick in the \emph{Kepler} bandpass. We show the cloud coverage for different cloud species (plain, colored lines) and for the cold trap model of Section~\ref{sec::CT} (dotted line). In the cold trap model, clouds of different composition disappear from the atmosphere when the equilibrium temperature is cooler than the following temperature: $1200\rm\,K$ (MnS), $1500\rm\,K$ (Cr), $1600\rm\,K$ ($\rm MgSiO_3$), $1700\,\rm K$ ($\rm Al_2O_3$) and $1800\rm\,K$ (Fe and $\rm CaTiO_3$). We used a particle size of $\SI{0.1}{\micro\meter}$ and three different cloud top pressure (rows).
}
 \label{fig::CloudCover}
\end{figure*}

Although observations indicate that a continuum from cloudy to cloudless limbs exist in the current hot Jupiter sample~\citep{Sing2016}, no correlations between the cloudiness at the limb and planetary parameters such as equilibrium temperature have been found, in contrast with the dayside cloud coverage discussed in this paper. 

We now use our model to determine the expected cloud fraction on the dayside, at the terminator and in the nightside with the assumption that only the thermal structure of the planet influences its cloudiness. An atmospheric column should be considered cloudy if at the photosphere most of the optical depth is due to cloud scattering rather than gas absorption and scattering. As a consequence, an atmospheric column should reflect more light if it contains clouds than if it does not. To calculate the cloud coverage we therefore compare the reflected flux maps described in Section~\ref{sec::Fluxes} with the flux maps obtained for a cloudless atmosphere. For the nightside case, we calculate additional flux maps by artificially illuminating the nightside of the planet as if it were the dayside\footnote{Note that the thermal structure used for the calculation is the one calculated by the global circulation model, assuming a non-illuminated nightside.}. We define a particular latitude and longitude point of the planet to be cloudy when the cloudy model reflects more flux than the cloudless one. 

 This method naturally takes into account the condensate abundance, the condensate opacity in the \emph{Kepler} bandpass, and the specific geometry both of the limb measurements and of the secondary eclipse measurements. Here we consider clouds formed by $\SI{0.1}{\micro\meter}$ size particles for which the scattering opacity dominates the extinction. Our calculations should therefore be relevant for the transit observations. 

Figure~\ref{fig::CloudCover} shows the cloud fraction on the dayside, the limb and the nightside as a function of equilibrium temperature for different species and different cloud top pressures. The cloud fraction always increases with decreasing temperature for a given cloud species, with the more refractory species covering a larger area of the planet. The cloud coverage depends on the cloud vertical extent: the higher the cloud extends, the greater the cloudiness. This effect is of first order for the limb and for the nightside but only of second order for the dayside cloudiness. This is a consequence of the strong irradiation received by the planet: it produces a shallower temperature gradient on the dayside than on the nightside and on the limb (see Figure~\ref{fig::PT} and Appendix). As a consequence, dayside clouds are more homogeneous in the vertical direction and thus less sensitive to a change in the cloud top pressure than nightside and terminator clouds. Note that the cloud coverage provided here is the area covered by optically thick clouds in the \emph{Kepler} bandpass. Although they are indicative of the cloud coverage, they depend on the ratio of the cloud scattering opacity to the gas extinction opacity. At wavelengths where this ratio is higher we expect a larger cloud coverage whereas at wavelengths where this ratio is smaller we expect a smaller cloud coverage.


For the cold trap model described in Section~\ref{sec::CT} and shown as a dotted black line in Figure~\ref{fig::CloudCover}, the cloud coverage is no longer a monotonic function of the equilibrium temperature. On the dayside, we expect most planets to be partially cloudy in agreement with the prediction that most of them should have a shift in their reflected lightcurve (see sec~\ref{sec::CT}). {\ct We predict a maximum cloudiness on the dayside for planets with an equilibrium temperature around $1900\,\rm K$, $1300\,\rm K$ and $1000\,\rm K$ and a minimum for planets with equilibrium temperatures around $1500\,\rm K$ and $1100\,\rm K$.} Interestingly, the hottest point of the dayside is almost always cloudless. This is a direct consequence of the irradiation. Irradiated atmospheric columns have a shallower pressure-temperature profile than non-irradiated columns (see Appendix). As a consequence, when the cloud deck hits the $\approx\SI{100}{\bar}$ level, the temperature in the hot spot is higher than the condensation temperature from $\approx 1-\SI{10}{\milli\bar}$ to $\SI{100}{\bar}$, leading to a cloudless hot spot. 

On the limb, the effective cloud coverage is a stronger function of the cloud vertical extent. Very high clouds ($P_{\rm top}=\SI{1}{\micro\bar}$) obscure the whole limb at any temperature: there is always a cloud species abundant enough to obscure the whole limb.{\ct~When clouds cannot form higher than $\SI{1}{\milli\bar}$, partial limb cloud coverage is expected in the $1400-1600\rm K$ equilibrium temperature range, as silicate clouds disappear from the atmosphere.} When clouds do not extend higher than $\SI{100}{\milli\bar}$ patchy clouds are expected over the limb of the planet for all planets, with a cloudiness that is of the order or less than $50\%$. Note that other minor species not considered here might become important at the limb because of the slant geometry and increase the cloud optical depth~\citep{Fortney2005b}. Another caveat is that the cloud coverage at the limb calculated here is determined by the temperature at the limb itself which is a region of small spatial extent where the temperature gradient is large and where multiple processes might further influence the thermal structure. Therefore approximations in the global circulation model (equilibrium chemistry, cloudless opacities, absence of drag) that affect the dayside structure to a second order might be of greater importance when determining the limb temperature. 

On the nightside, the cloud coverage is higher than on the dayside and on the terminator. In most cases the nightside of the planet is always entirely covered by clouds. Only when clouds do not extend at pressures lower than $\SI{100}{\milli\bar}$ the nightside appears partially cloudy.

The cloud coverage calculated here is based solely on the thermal structure of the planet. Vertical mixing is taken into account through the cloud top pressure parameter but large scale, three-dimensional mixing in hot Jupiter can also produce horizontal variations in the cloud coverage. As an example a cloudless equator and cloudy poles are predicted by \citet{Parmentier2013} for HD209458b. The thermal and the dynamical effects can be seen as independent mechanisms shaping the cloudiness of the planet, the total cloudiness being a combination of both. Any horizontal structure in the cloudiness due to dynamical mixing should decrease the apparent cloudiness and the cloudiness derived in Figure~\ref{fig::CloudCover} should be seen as a maximum value. 

We show that partially cloudy atmospheres are expected over a very large range of equilibrium temperatures. Currently, most models retrieving the thermal structure and abundances from secondary eclipses, phase curves and transmission spectra use a one dimensional model, where the cloudiness is constant over the planet. When interpreting observations of a partially cloudy planet, however, this can lead to biased conclusions~\citep{Line2016} and artificially small uncertainties in the retrieved abundances. Our work points towards the necessity for more complex, partially cloudy models, as has been used in the brown dwarf community~\citep{Marley2010}.

We predict that some level of clouds are always present at the terminator of hot Jupiters, even for very high temperature planets. As the temperature contrast on the dayside increases with increasing equilibrium temperature, the western limb is always cold and cloudy. This provides an explanation for the presence of strong cloud signatures in the transmission spectrum of very high equilibrium temperature planets such as WASP-12 b~\citep{Sing2013}. 

{The cloud fraction can be used as a probe of the cloud composition. For example, the measurement of a partial cloud coverage at the limb on a planet with an equilibrium temperature of $1200\,\rm K$ such as HD189733b would point towards a cloud composition made of sodium sulfide clouds rather than more refractory material such as manganese sulfide clouds. The cloud coverage could be measured with high enough signal-to-noise ratio transmission spectrum with HST/WFC3 or JWST~\citep[see][]{Line2016}}.

We predict that the cloud coverage at the limb is more strongly affected by the vertical extent of the cloud than by the change of the thermal structure of the planet with equilibrium temperature. Trends relating equilibrium temperature and the importance of clouds over this range of equilibrium temperature~\citep[e.g.][]{Stevenson2016} might therefore be due to a variation in cloud top level rather than a variation in the thermal structure of the atmosphere.
 
We predict that partially cloudy daysides are common but that the hottest point of the dayside should almost always be devoid of clouds. As a consequence, secondary eclipse spectra should be even more dominated by the emission from the hottest point of the dayside than in a cloudless scenario.

Our models also show that clouds are always present on the nightside of hot Jupiters, providing a possible explanation for the low flux observed by HST/WFC3 from the nightside of Wasp-43b~\citep{Stevenson2014b}, as suggested by~\citet{Kataria2015}.

\section{Conclusion}

An asymmetry in the Kepler lightcurve has been observed for a handful of planets: for the hottest planets, the lightcurve peaks before secondary eclipse, whereas for planets cooler than $\sim1900\rm\,K$, it peaks after secondary eclipse~\citep{Esteves2015}. Here we model the thermal structure of tidally locked giant exoplanets with a state-of-the-art global circulation model. Based on the thermal structure we calculate the expected cloud distribution and planetary lightcurve in the \emph{Kepler} bandpass for different cloud physical and chemical properties and compare our results to the observations. {\ct Our main findings are summarized in Table~\ref{table::conclusions}. Predictions from our models can be found in Table~\ref{table::predictions}.}

\begin{deluxetable*}{ll}
\tablecaption{ Main conclusions and corresponding observational evidence \label{table::conclusions}} 
\tablehead{
\colhead{Conclusion}  & \colhead{Evidence} 
}
\startdata
Presence of silicate clouds for $T_{\rm eq}>1600\rm\,K$ & Large Kepler lightcurve offsets of Kepler-41b, Kepler-8b and Kepler-7b\\
								& Small albedo for planets with $1700\rm\, K<T_{\rm eq}<1900\rm\, K$ and large albedo for Kepler-7b\\
\\
Lack of silicate clouds for $T_{\rm eq}<1600\rm\,K$&  Large Kepler lightcurve offset of Kepler-12b\\
						    &Presence of three low albedo planets at $T_{\rm eq}<1600\,\rm K$\\
						    & Lack of high albedo planets at $T_{\rm eq}<1600\,\rm K$\\
						     \\
Presence of MnS clouds for $T_{\rm eq}<1600\rm\,K$ & Large Kepler lightcurve offset of Kepler-12b \\
\\
Lack of corundum and iron clouds for $T_{\rm eq}<1900\rm\,K$  &  Offsets in the Kepler lightcurve of Kepler-41b, Kepler-8b, Kepler-7b and Kepler-12b\\  				   
\\
Presence of gaseous TiO for $T_{\rm eq}>1900\rm\,K$  & Small offsets and large apparent albedos of Kepler-76b and HAT-P-7b\\
\\
\enddata
\end{deluxetable*} 
We show that the lightcurve of a planet is the sum of a thermal component and a reflected component. The thermal lightcurve is determined by the thermal structure of the planet and always peaks before the secondary eclipse of the planet due to the eastward displacement of the hot spot. The reflected lightcurve is determined by the cloud distribution and always peaks after the secondary eclipse as clouds tend to form west of the substellar point, where the atmosphere is cooler. We predict that a change from a reflected to a thermal dominated lightcurve should occur at an equilibrium temperature of $\sim1900\,\rm K$, naturally explaining the trend from negative to positive offset in the \emph{Kepler} data.

For the cooler planets, we show that the presence of an asymmetry of atmospheric origin in the lightcurves observed by the \emph{Kepler} spacecraft is a telltale sign of the cloud chemical composition, independently of the particle size and of the cloud top pressure. Among several important cloud species, we show that only silicates and manganese sulfide clouds can produce a noticeable offset in the \emph{Kepler} lightcurve. Models incorporating silicate clouds predict a large offset of the \emph{Kepler} lightcurve in the $1600-1900\rm\,K$ equilibrium temperature range. They provide a good match of the phasecurve offset and albedos of Kepler-7b, Kepler-8b and Kepler-41b. However, they predict a high albedo and no offset for the cooler planet Kepler-12b, in stark contrast with the observations. Conversely, models considering manganese sulfide clouds predict a large offset for equilibrium temperatures between $1300$ and $1700\rm\,K$. They provide a good match for the observations of Kepler-12b, Kepler-7b and Kepler-8b but cannot reproduce the large offset observed for Kepler-41b. 

We suggest that a transition between silicates and manganese sulfide clouds happens around $T_{\rm eq}\approx1600\,\rm K$. The rainout of the cloud species as they are cold trapped in the deep layers of the atmosphere provides a natural explanation for this transition. This is analogous to the L/T brown dwarf transition where silicate clouds disappear as they are cold trapped bellow the photosphere when the equilibrium temperature reaches $\approx1400\,\rm K$~\citep{Kirkpatrick2005}. We expect a similar transition to happen for other cloud species, leading to a temperature-dependent cloud composition for hot Jupiters. {\ct As a consequence we do not expect silicate clouds to be present in the canonical hot Jupiter HD189733b ($T_{\rm eq}=1200\,\rm K$) or in the hotter HD209458b ($T_{\rm eq}=1450\,\rm K$) as has often been proposed~\citep[e.g.][]{LecavelierDesEtangs2008}. Instead our models imply that manganese sulfide clouds should be the dominant cloud species in both planets (in agreement with the albedo spectra of HD189733b, see~\citet{Barstow2014}), with the possible addition of chromium clouds in HD209458b and sodium sulfide clouds in HD189733b.}
\begin{deluxetable*}{ll}
\tablecaption{ Main predictions and corresponding observational tests \label{table::predictions}} 
\tablehead{
\colhead{Predictions}  & \colhead{Observational test}
}
\startdata
Variation of the cloud composition with $T_{\rm eq}$ & Detect the condensate spectral signature in the far IR transit spectrum\\
\\
Geometric albedo variation with $T_{\rm eq}$ & Measure secondary eclipses in the optical \\
 \\
Partially cloudy limbs & Detect the shape of molecular features in the transmission spectrum\\
\\
Cloudy nightsides & Measure the thermal lightcurves\\
\\
Cloudless dayside hot spots & Detect molecular features in the secondary eclipse infrared spectrum\\
&Detect phase shifts in planetary optical phase curves
\\
\enddata
\end{deluxetable*} 

{\ct The disappearance of the silicate clouds due to the deep cold trap places a combined constraint on the deep vertical mixing rate at the 10-100 bar cold trap, and on the particle size at that cold trap as the mixing rate must be small and/or the particle size must be large for the cold trap to be efficient~\citep{Spiegel2009,Parmentier2013}. This observational constraint is consistent with the theoretical expectation that cloud particles should increase in size with increasing pressure~\citep{Lee2015} and that the vertical mixing rate should decrease with increasing pressure in the radiative atmosphere of hot Jupiters~\citep{Parmentier2013}.}

With our cold-trap model we reproduce all current observed phase shift offsets and albedos in the \emph{Kepler} bandpass. We predict that phase shifts in the reflected lightcurve of hot Jupiters should be common over the $1400-1800\,\rm K$ equilibrium temperature range. We also predict an increase in the planet geometric albedo in the \emph{Kepler} bandpass for planets with an equilibrium temperature between $1600$ and $1700\,\rm K$ due to the presence of silicate clouds and with an equilibrium temperature around $1200-1300\,\rm K$ due to the presence of manganese sulfide clouds, making them easier targets for secondary eclipse observations at small wavelengths. 

We also show that the albedo and offset of planets cooler than $2000\,\rm K$ are well explained by models without TiO/VO whereas the presence of TiO/VO allows a better fit of the \emph{Kepler} lightcurves offset and apparent albedo of the hottest planets Kepler-76b and HAT-P-7b. Such a transition for TiO/VO is expected from our cold trap model and is compatible with current detections of TiO in hot Jupiter atmospheres.

Finally, we show that the temperature contrast in the dayside atmosphere increases with equilibrium temperature, leading to an always cold western limb, naturally explaining why clouds are present at the limb of a large number of hot Jupiters atmospheres, even in the hottest ones. We highlight the fundamental three-dimensional structure of hot Jupiters atmospheres. Nightsides are predicted to be always cloudy, providing a possible explanation for the low flux observed in the nightside of some planets. Inhomogeneous limbs and daysides are expected and should affect the retrieval of molecular abundances from the transmission spectra whereas the dayside spectrum should often be dominated by a cloud-free hot spot. 

{\ct Observations of the apparent albedo in the optical and of the phase shift of the optical phase curve that will be obtained by future photometric space mission such as CHEOPS, TESS and PLATO should be able to confirm or refute the presence of temperature dependent cloud composition. Particularly, observations of planets albedos in the $1500-1700\,\rm K$ equilibrium temperature range should be able to confirm the disappearance of silicate clouds. 
 
Transit spectroscopy can also provide insights on the cloud composition by measuring the cloud coverage at the limb of the planet or by directly measuring the spectral signature of the cloud species~\citep{Wakeford2015}. Higher signal-to-noise HST/WFC3 spectrum and future JWST observations should be able to distinguish between different cloud composition for a given planet, providing important insights on the deep temperature profile and the mechanisms governing the long-term evolution of hot Jupiters.}

\section*{ACKNOWLEDGEMENTS}
We thank Mike Line for reading the manuscript and providing useful comments and Kevin Stevenson for useful discussions. V.P. acknowledges support from the Sagan Postdoctoral Fellowship through the NASA Exoplanet Science Institute. A.P.S. was supported by Origins grant NNX12AI196.

 \bibliography{Parmentier2015-Submitted.bbl}
\appendix
\section{Pressure-temperature profiles}
We show in Figures~\ref{fig::PTprofiles1} to~\ref{fig::PTprofiles3} the pressure-temperature profiles of a sample of our grid of models for planets with an equilibrium temperature from $1300\,\rm K$ to $2200\,\rm K$ together with the condensation curves of several important species. In the cold trap model of Section~\ref{sec::CT}, we consider that the species rain out of the atmosphere and remove them from the calculation when their condensation curves crosses the pressure-temperature profile in the $10-\SI{100}{\bar}$ region. We consider that the same mechanism is at work for titanium dioxide and provide additional models where it is present in solar composition in Figure~\ref{fig::PTprofiles2}. Given the long radiative timescales at these large depths and the limited time for which the simulations ran, the deep temperature is strongly influenced by our initial condition. Our initial condition is the global average temperature profile calculated by~\cite{Parmentier2015}, which is obtained by fitting the analytical formula of ~\cite{Parmentier2015} to the global average, solar-composition, without TiO/VO, cloudless profiles calculated by ~\cite{Fortney2007}. Although meridional and vertical transport of heat should affect this global average on the long-term, these variations are expected to be small compared to the variations due to other uncertainties of the problem such as the planet chemical composition~\citep[e.g. figure 10 of][]{Heng2011a}.

 \begin{figure}
   \begin{minipage}[c]{.46\linewidth}
\includegraphics[width=\linewidth]{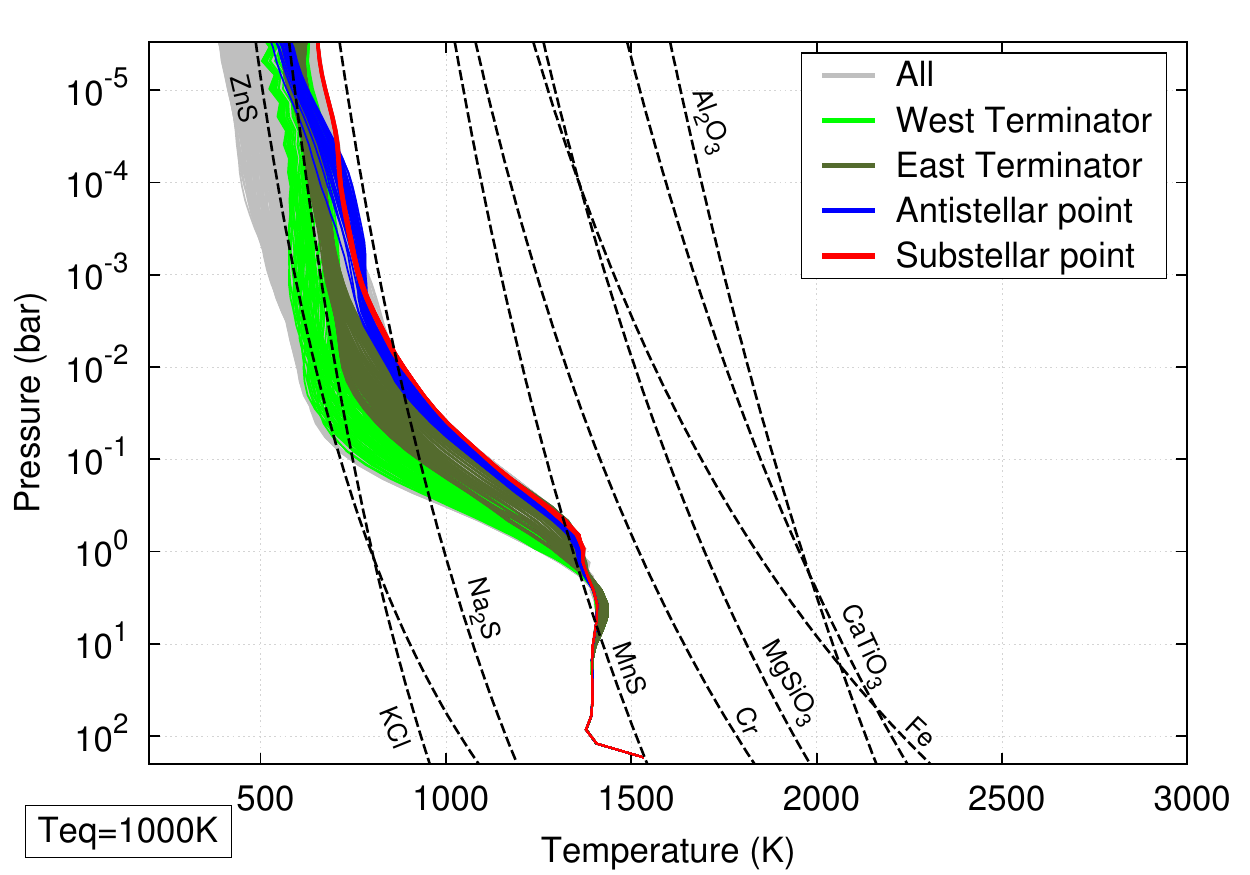}   
\includegraphics[width=\linewidth]{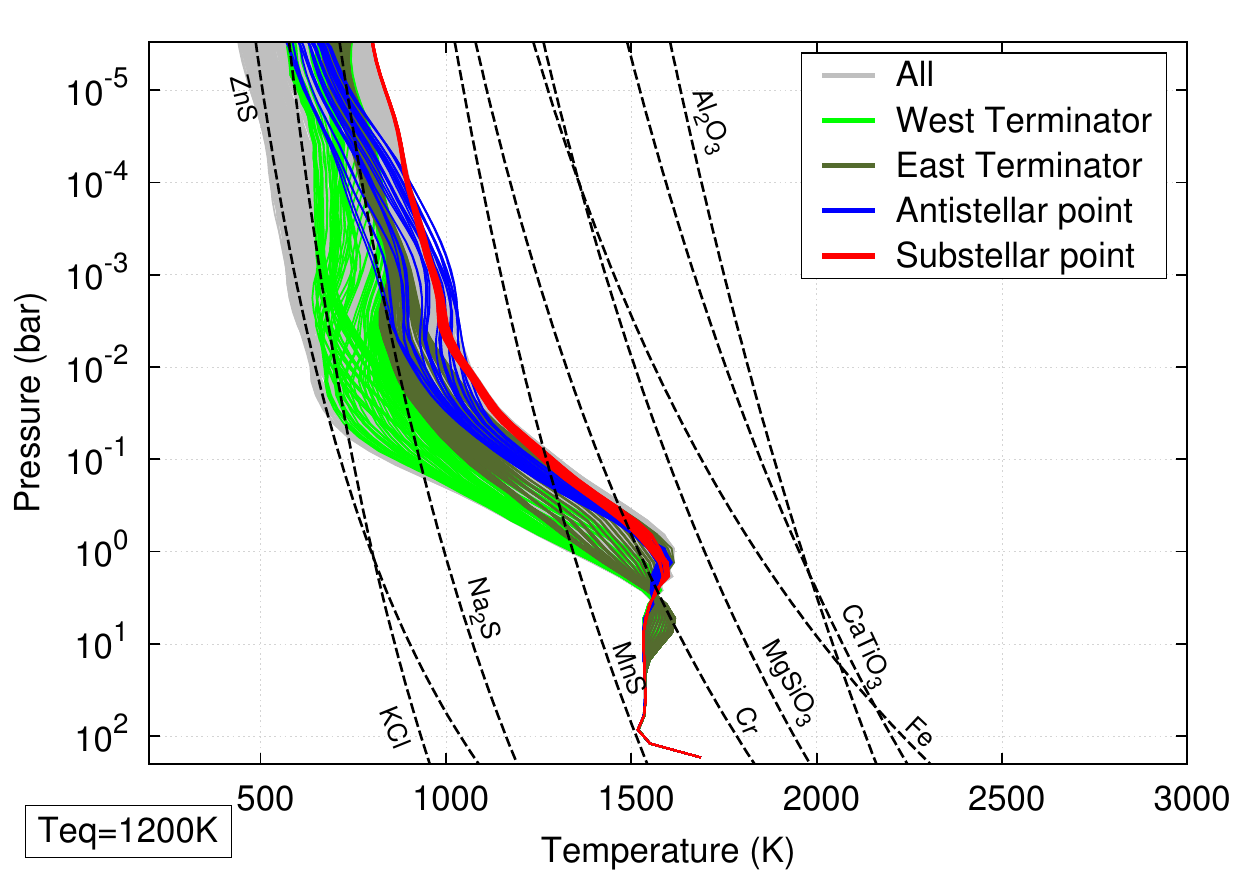}
\includegraphics[width=\linewidth]{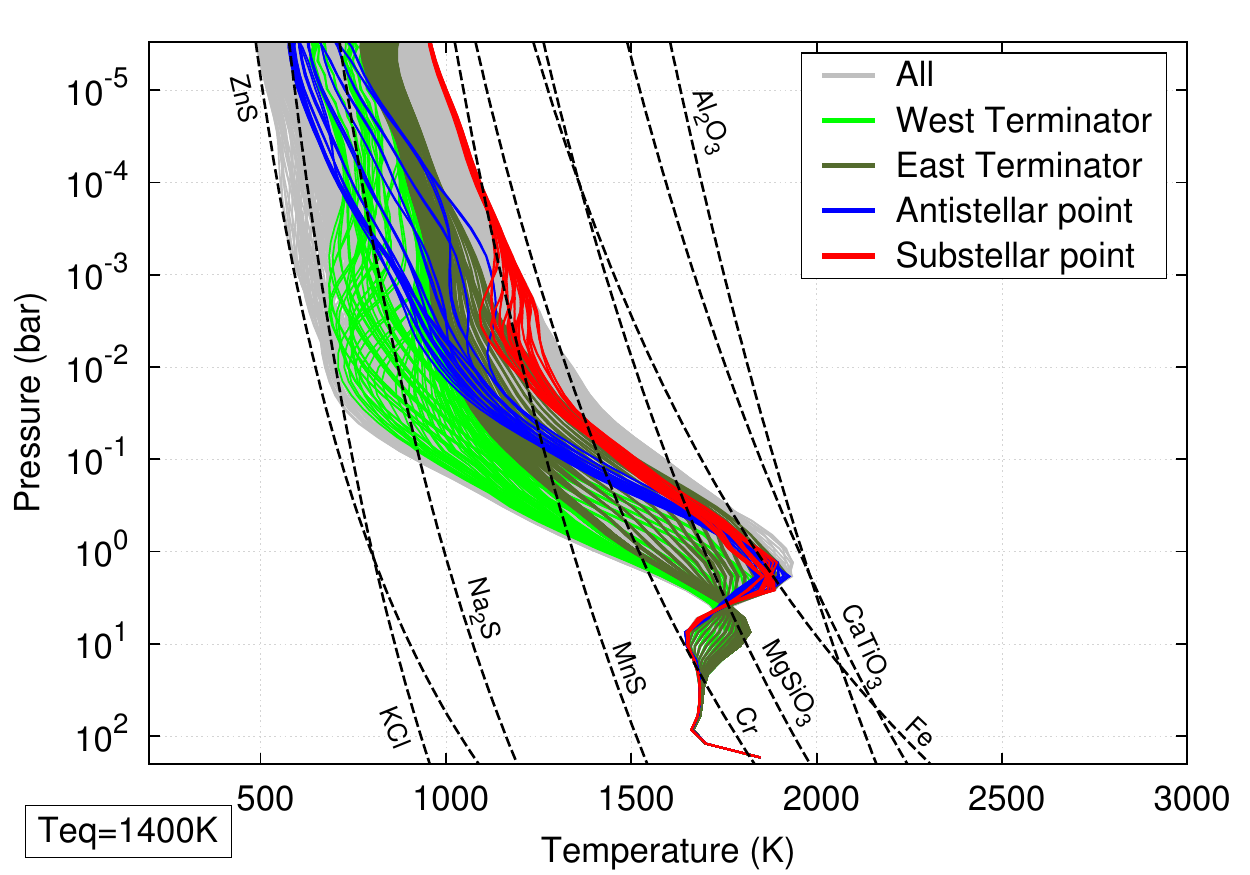}
   \end{minipage} \hfill
   \begin{minipage}[c]{.46\linewidth}
\includegraphics[width=\linewidth]{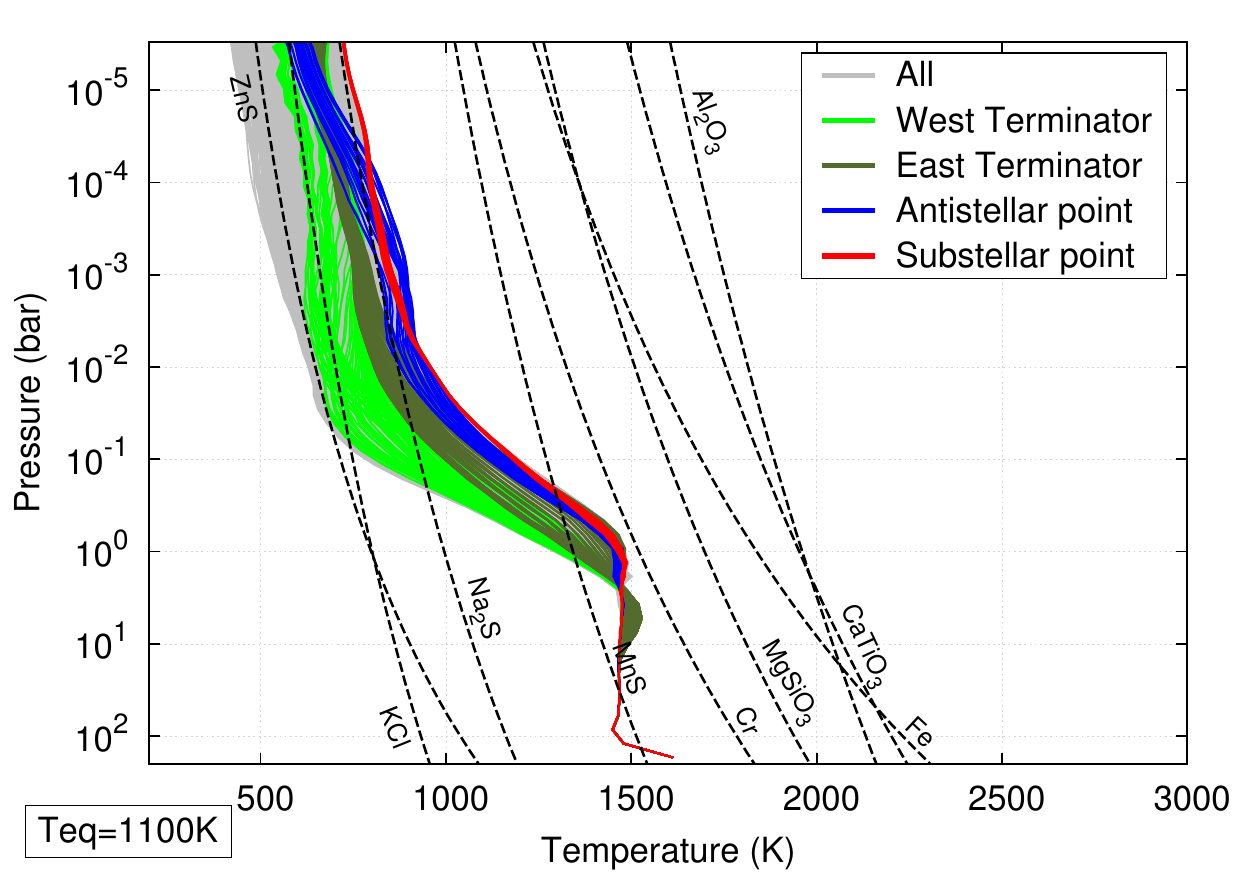}
\includegraphics[width=\linewidth]{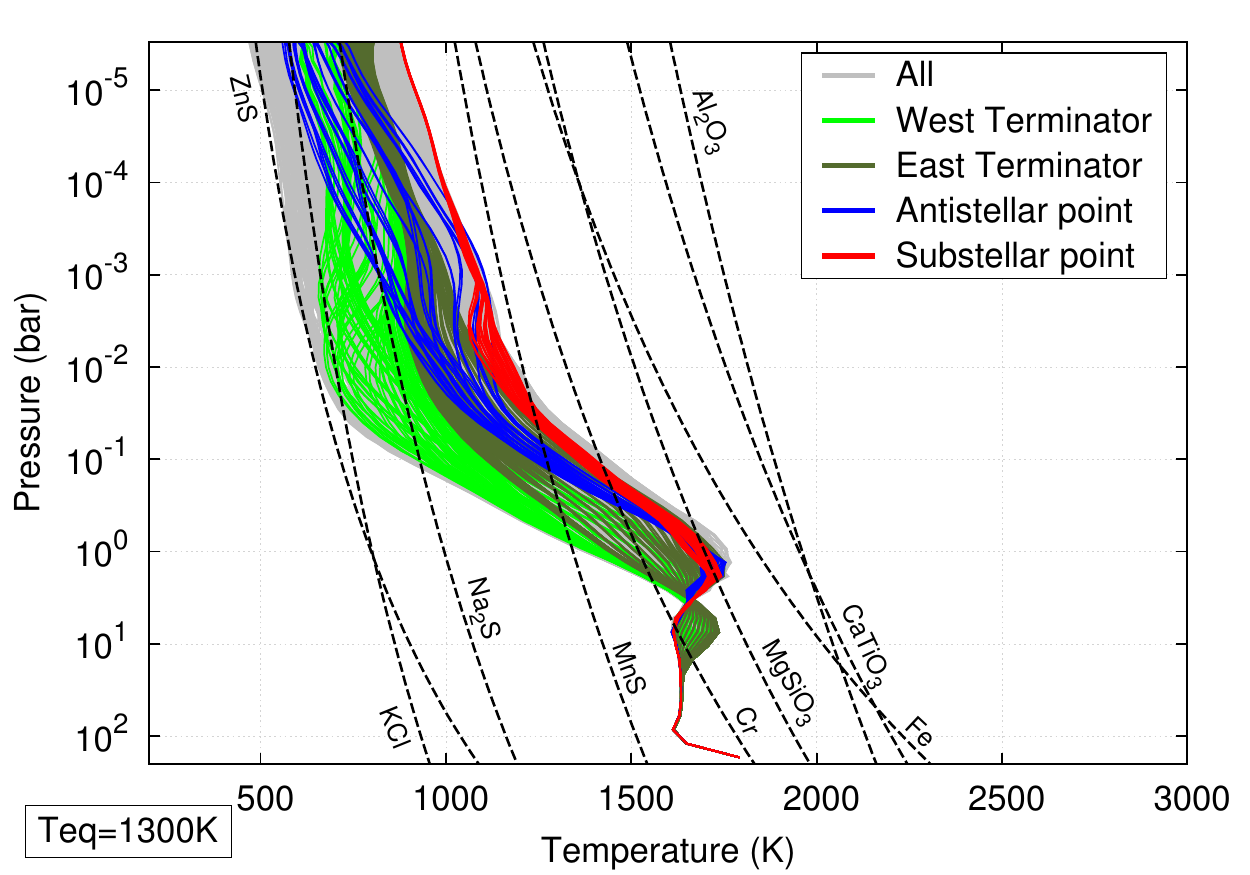}
\includegraphics[width=\linewidth]{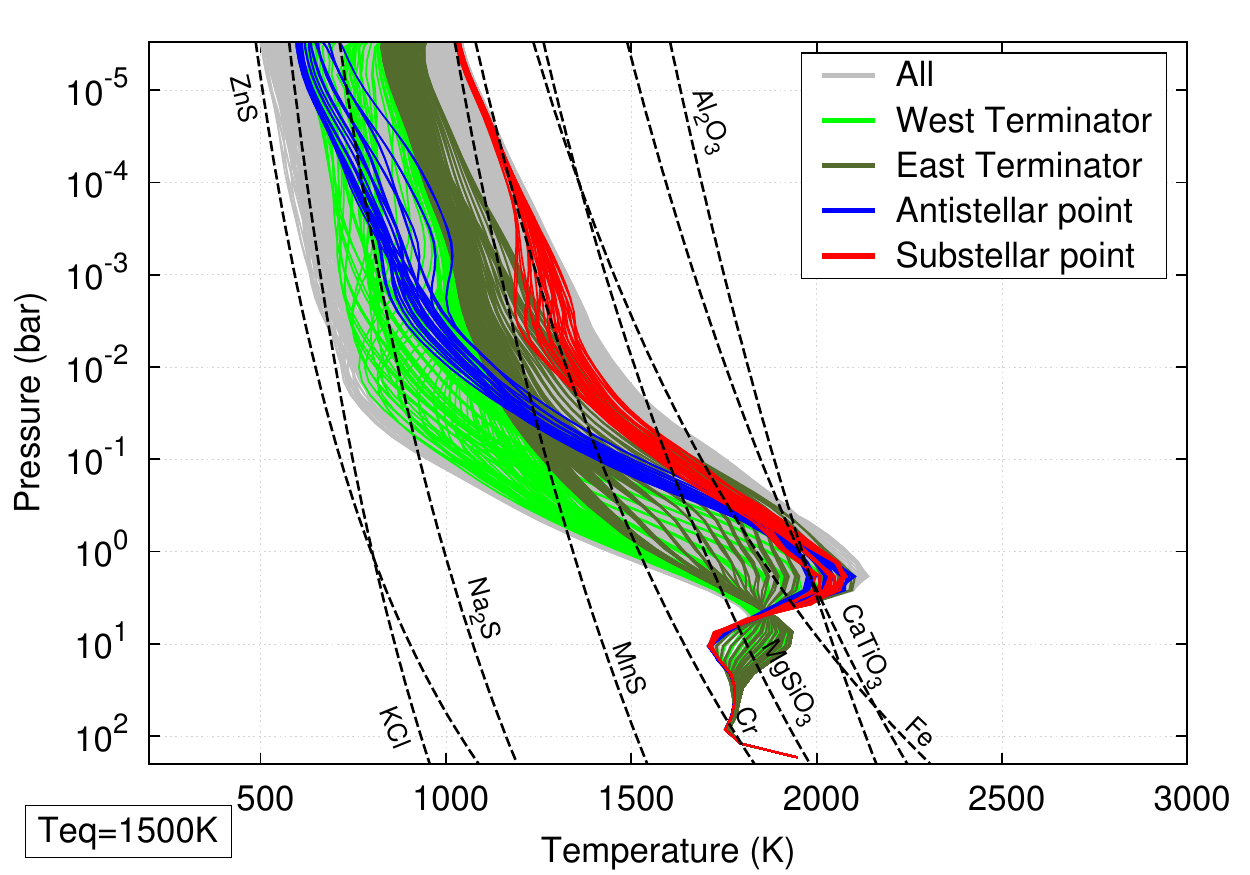}

 \end{minipage}
 \caption{Pressure temperature profiles for our grid of GCM models without TiO/VO. The profiles are colored with respect to their spatial position as in Figure~\ref{fig::PT}. Superposed are the condensation curves of the species considered in this paper.}
   \label{fig::PTprofiles1}
\end{figure}

 \begin{figure}
   \begin{minipage}[c]{.46\linewidth}
\includegraphics[width=\linewidth]{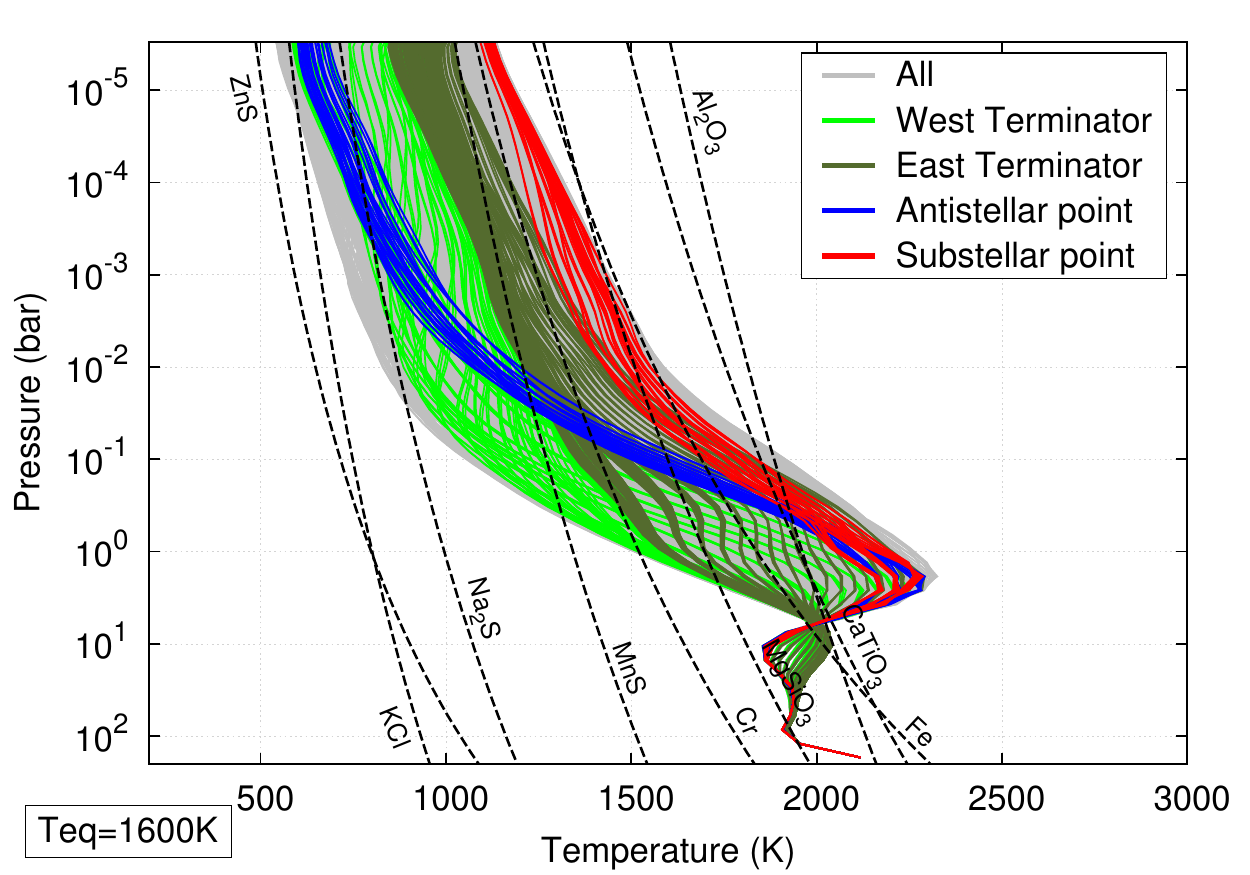}
\includegraphics[width=\linewidth]{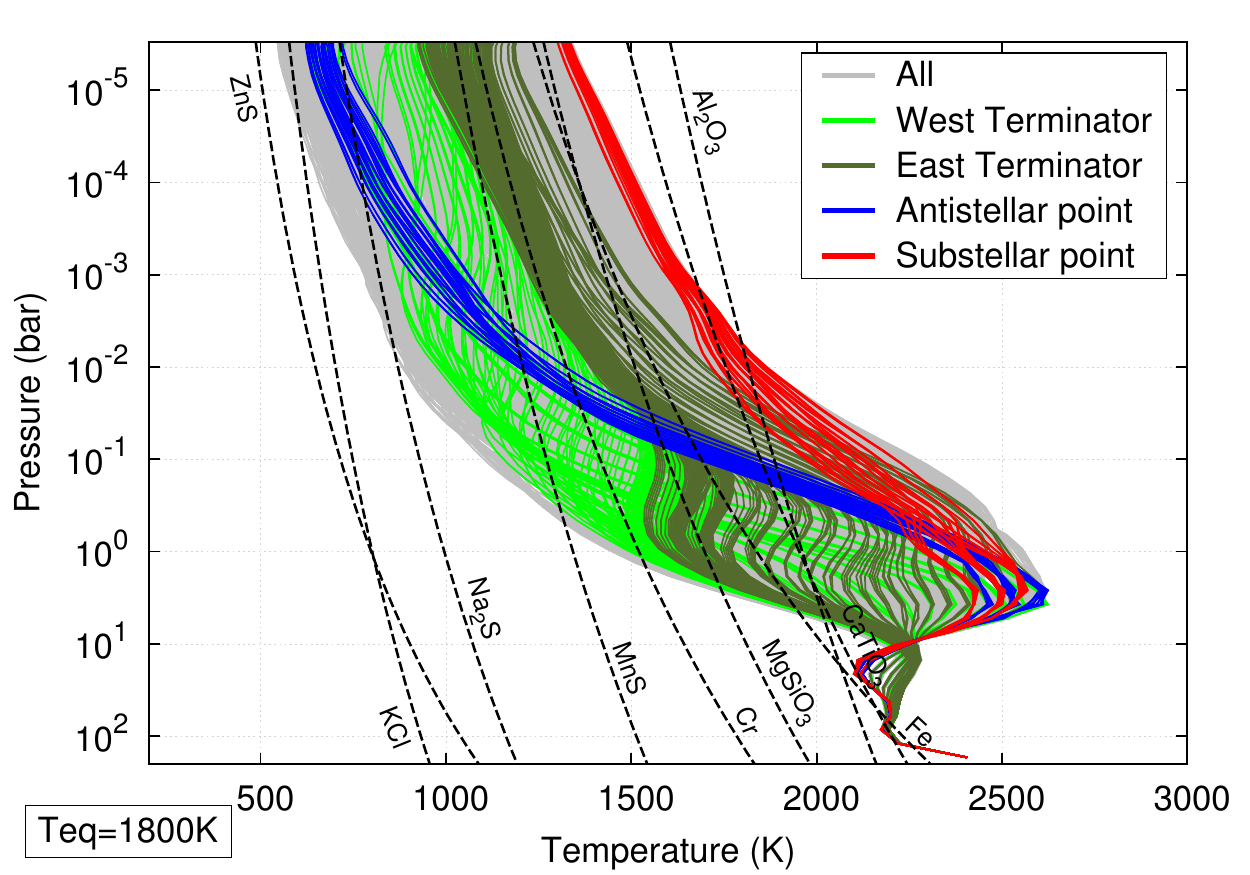}
   \end{minipage} \hfill
   \begin{minipage}[c]{.46\linewidth}
\includegraphics[width=\linewidth]{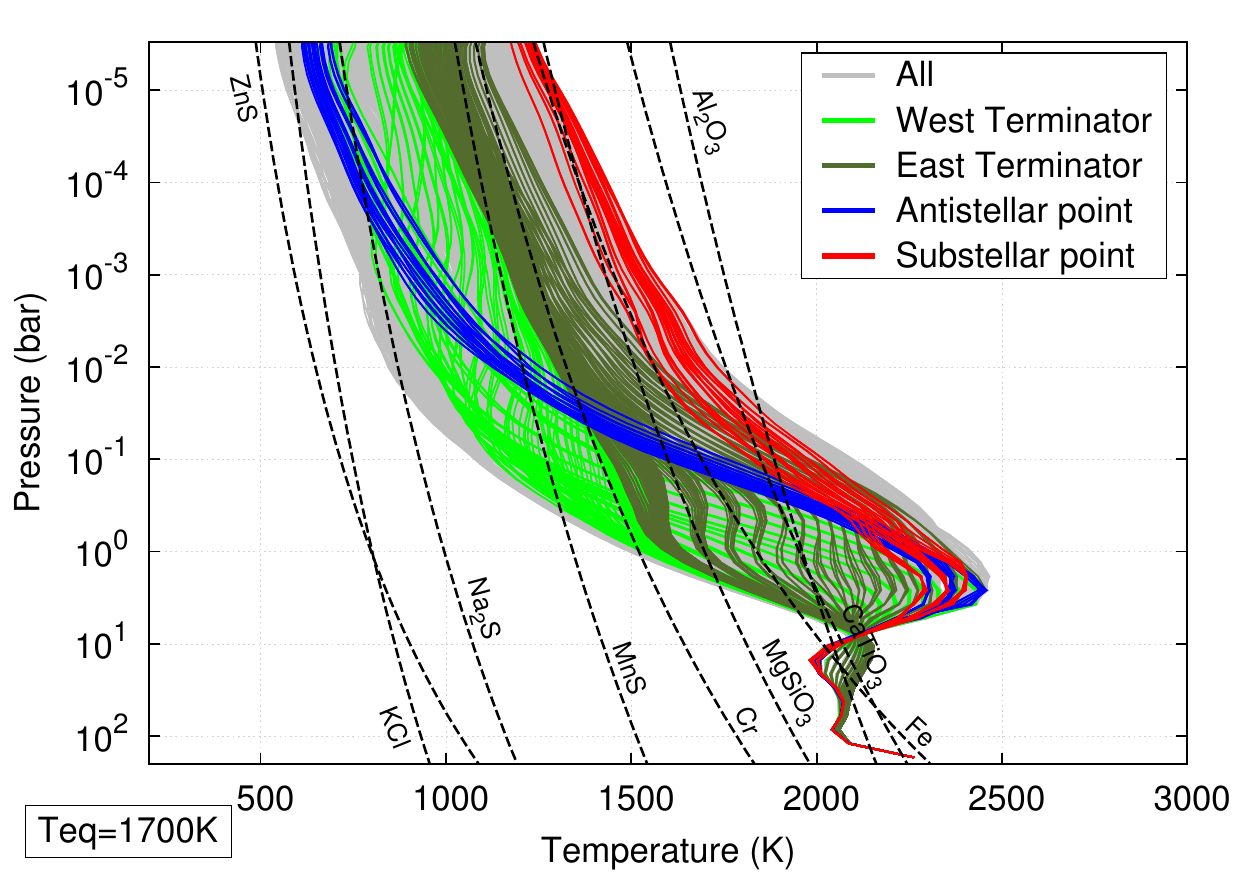}
\includegraphics[width=\linewidth]{./Temp-Teq1900-Spag-2Limb.pdf}
 \end{minipage}
 \caption{Pressure temperature profiles for our grid of GCM models without TiO/VO. The profiles are colored with respect to their spatial position as in Figure~\ref{fig::PT}. Superposed are the condensation curves of the species considered in this paper.}
   \label{fig::PTprofiles2}
\end{figure}

 \begin{figure}
   \begin{minipage}[c]{.46\linewidth}
\includegraphics[width=\linewidth]{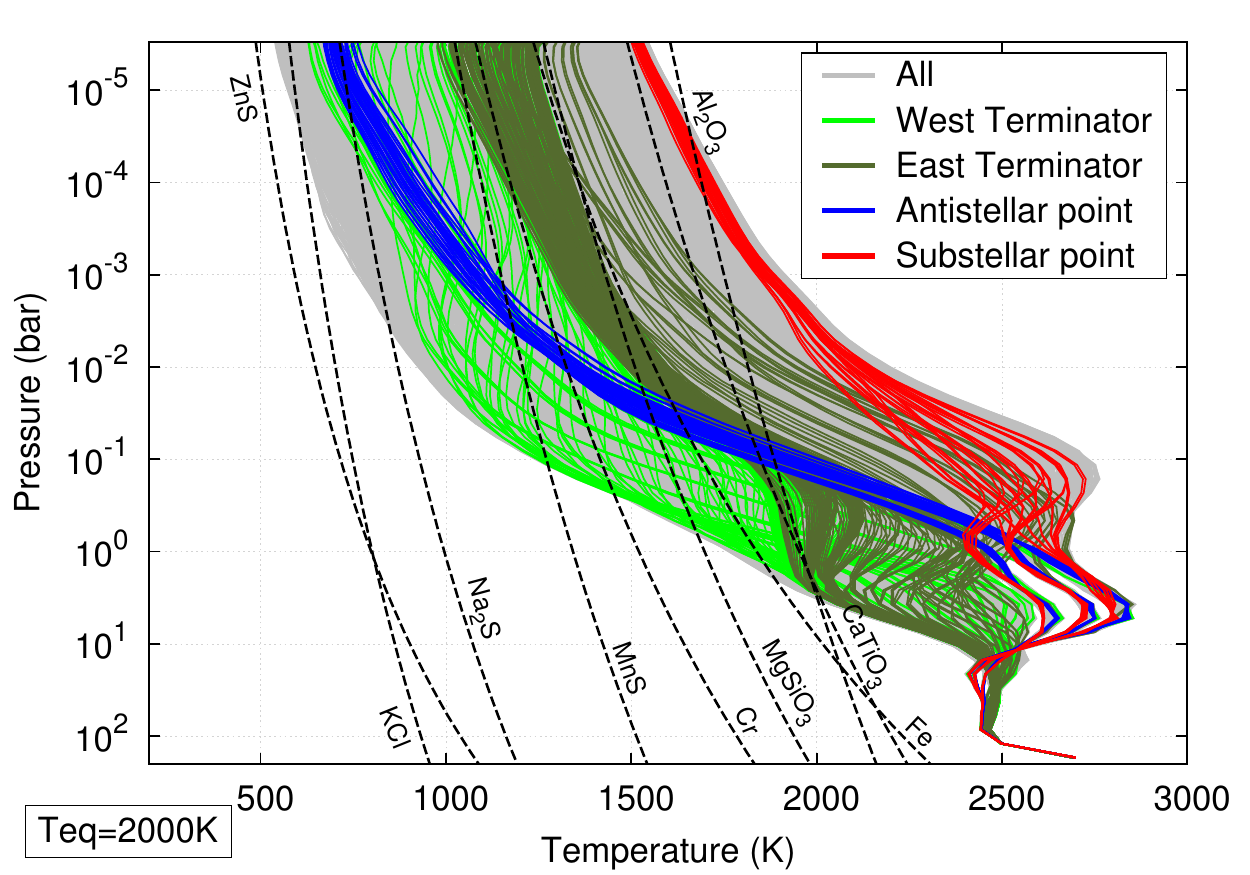}
\includegraphics[width=\linewidth]{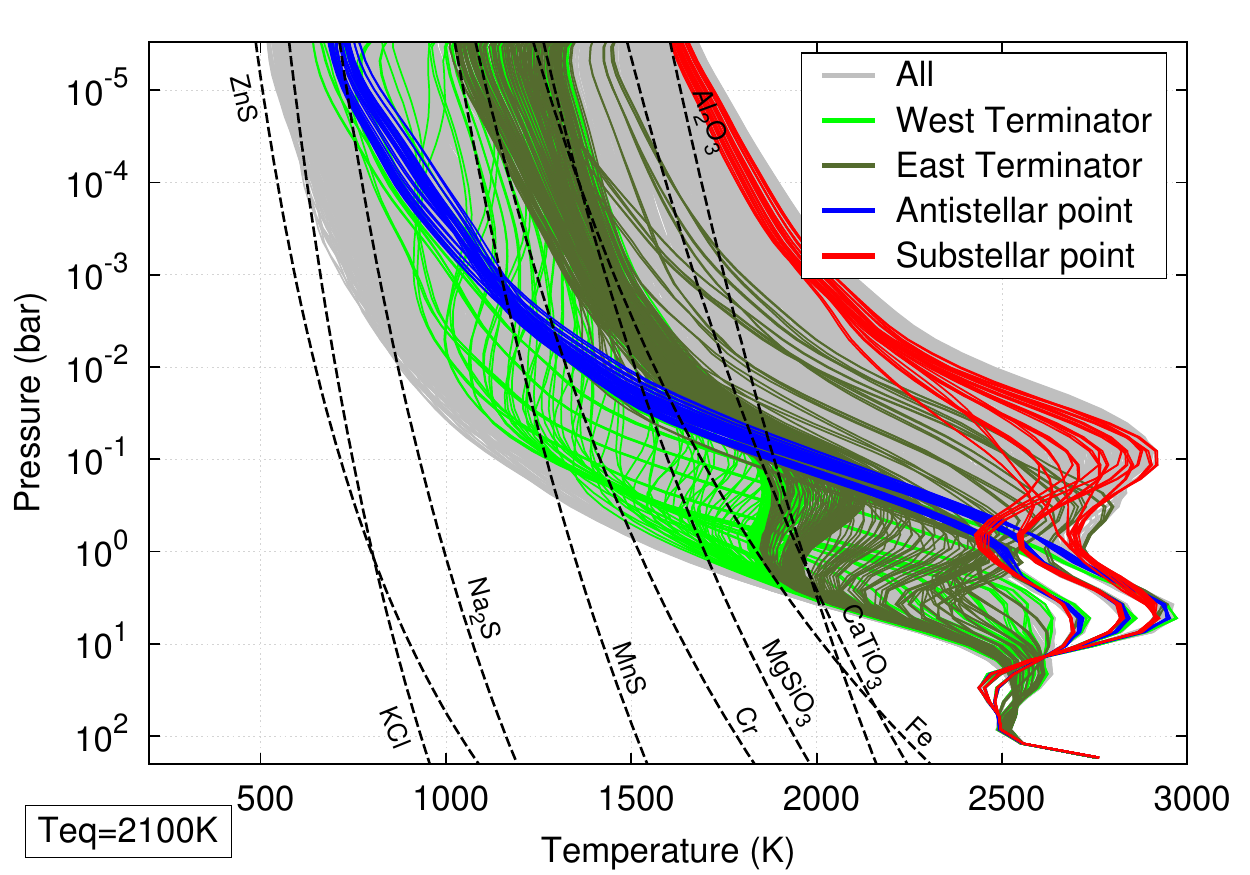}
\includegraphics[width=\linewidth]{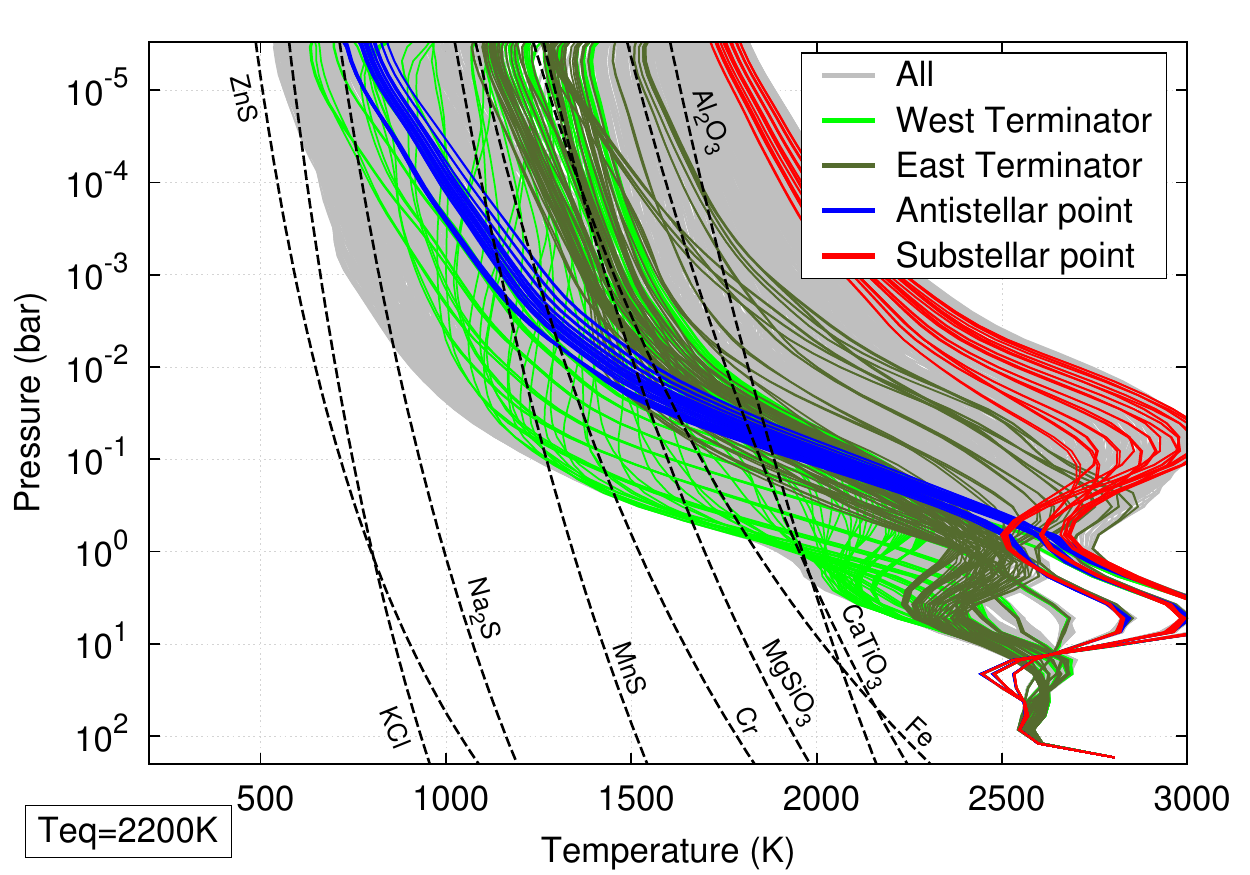}

   \end{minipage} \hfill
   \begin{minipage}[c]{.46\linewidth}
\includegraphics[width=\linewidth]{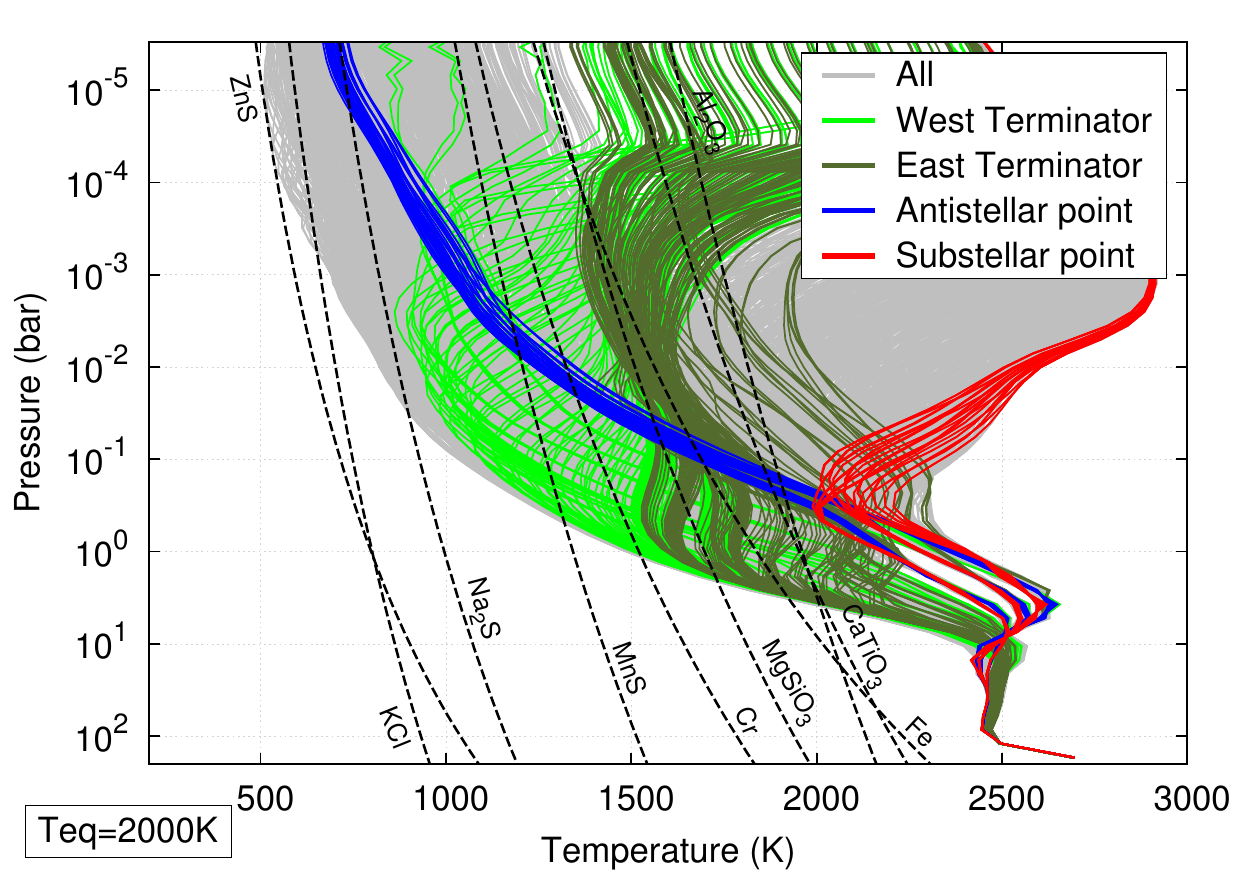}
\includegraphics[width=\linewidth]{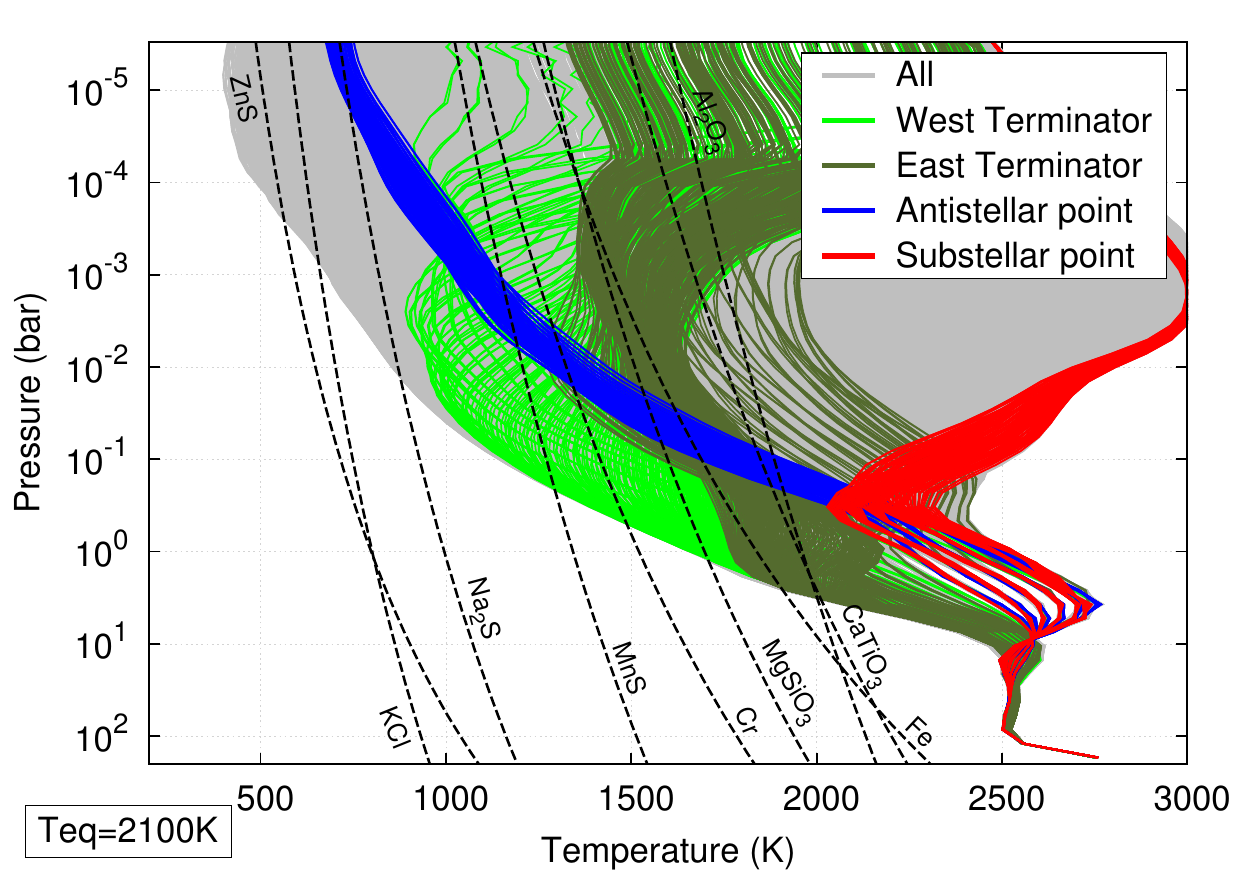}
\includegraphics[width=\linewidth]{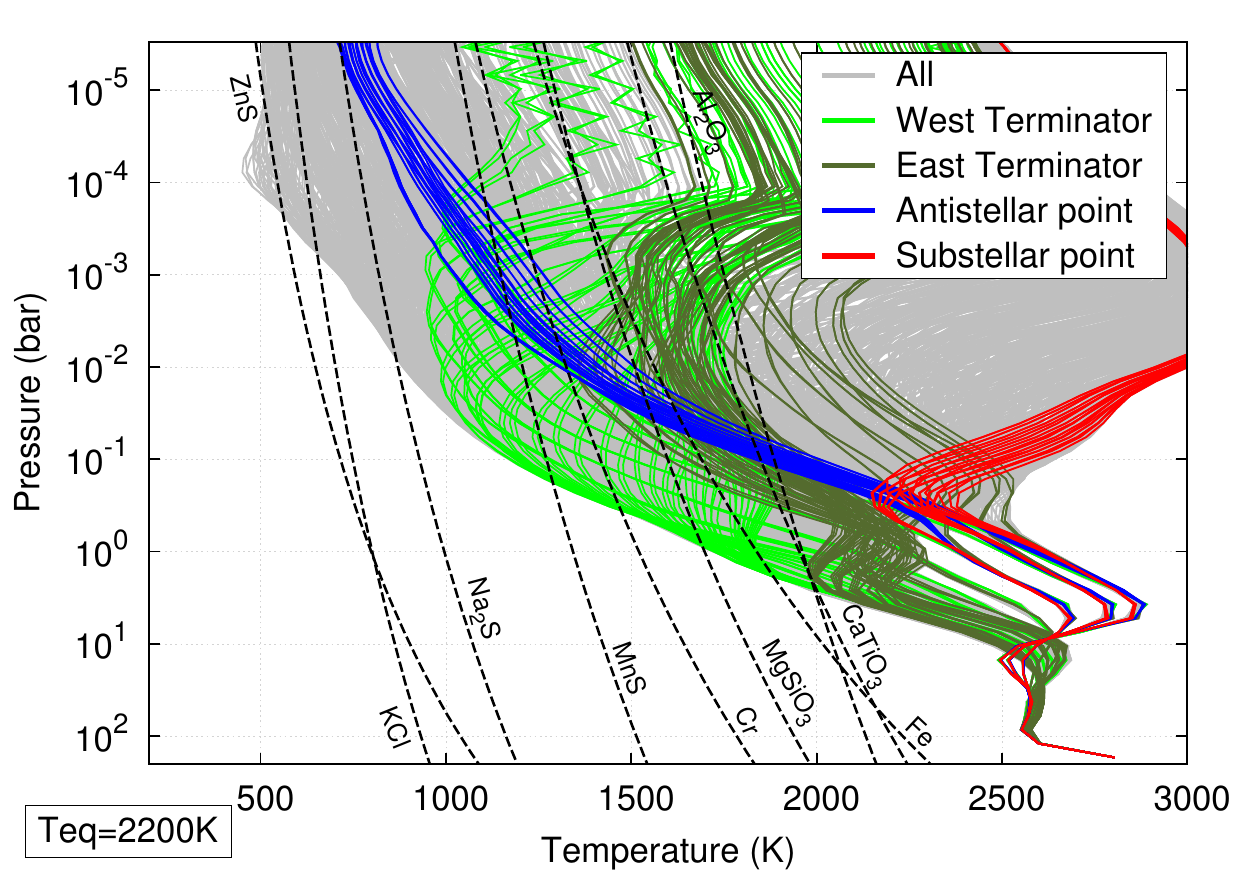}

 \end{minipage}
 \caption{Pressure temperature profiles for our grid of GCM models without TiO/VO (left) and with TiO/VO (right). The profiles are colored with respect to their spatial position as in Figure~\ref{fig::PT}. Superposed are the condensation curves of the species considered in this paper.}
   \label{fig::PTprofiles3}
\end{figure}

\section{Optical depths}
{\ct We show in Figure~\ref{fig::opd} the integrated extinction optical depth at 1bar for different cloud species in the dayside of our grid of hot Jupiter models. Clouds are present only at the limb at high equilibrium temperature and cover homogeneously the dayside of the planet at low equilibrium temperatures. The equilibrium temperature for which partial clouds are present is a strong function of the cloud composition. When several clouds are present at the same time, the figure can be used to compare the relative importance of the different clouds.  

The calculated optical depths are larger than the ones calculated~\citep{Marley2000} using a similar model for brown dwarf atmospheres due to the lower gravity of hot Jupiters. They are also much larger than the ones estimated by~\citet{Fortney2005} for HD209458b.~\citet{Fortney2005} used the~\citet{Ackerman2001} model with a rather small vertical mixing coefficient, leading to a vertically thin cloud formed of large ($\approx10-\SI{100}{\micro\meter}$) particles. Our model lacks of the self-consistent structure of the~\citet{Ackerman2001}. However, the small particles and vertically homogeneous cloud we assume, corresponding to a large vertical mixing coefficient, are more in agreement with transmission spectrum observations and theoretical work on vertical mixing~\citet{Parmentier2013}. Overall, the calculations of Figure~\ref{fig::opd} should be seen as the maximum optical depth that can be reached by clouds in hot Jupiter atmospheres, as larger particles and weaker mixing rates would tend to decrease the opacities.}

\begin{sidewaysfigure}
\includegraphics[width=\linewidth]{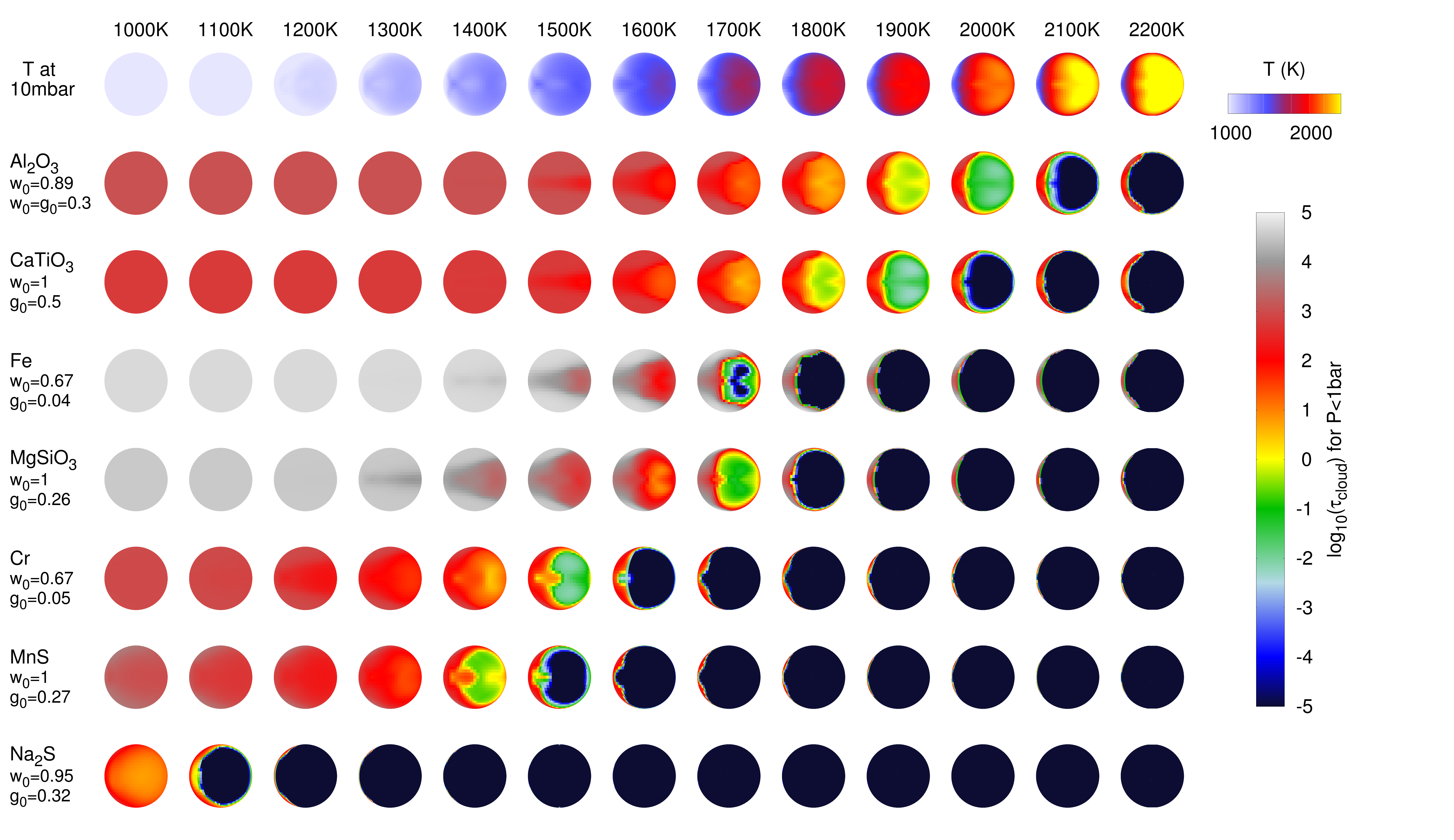}
 \caption{Integrated vertical extinction cloud optical depth for P<1bar calculated in the Kepler bandpass for different cloud composition (rows) and planet equilibrium temperatures (columns). Here we assume a particle size of $\SI{0.1}{\micro\meter}$ and a cloud top pressure of $\SI{1}{\micro\bar}$. When the optical depth is lower than 1, the clouds are transparent.}
 \label{fig::opd}
\end{sidewaysfigure}
\end{document}